%% file: main.tex
\documentclass[AMA,STIX1COL]{WileyNJD-v2}

\articletype{Article Type}%

\received{-}
\revised{-}
\accepted{-}

\usepackage{mathtools, upgreek}

\DeclareMathOperator{\dd}{d\hspace{-2pt}}
\DeclareMathOperator{\Pe}{Pe}

\begin{document}

\title{A phase-field model for active contractile surfaces}
\author[1,2]{Sebastian Aland*}
\author[1]{Claudia Wohlgemuth}

\address[1]{\orgdiv{Institute of Numerical Mathematics and Optimization}, \orgname{TU  Freiberg}, \orgaddress{Akademiestaße 6, 09599 Freiberg, \country{Germany}}}
\address[2]{\orgdiv{Faculty of Computer Science/Mathematics}, \orgname{HTW Dresden}, \orgaddress{Friedrich-List-Platz 1, 01069 Dresden, \country{Germany}}}

\corres{*Sebastian Aland \email{sebastian.aland@htw-dresden.de}}


\abstract[Summary]{
The morphogenesis of cells and tissues involves an interplay between chemical signals and active forces on their surrounding surface layers. The complex interaction of hydrodynamics and material flows on such active surfaces leads to pattern formation and shape dynamics which can involve topological transitions, for example during cell division. 
To better understand such processes requires novel numerical tools. Here, we present a phase-field model for an active deformable surface interacting with the surrounding fluids. The model couples hydrodynamics in the bulk to viscous flow along the diffuse surface, driven by active contraction of a surface species. As a new feature in phase-field modeling, we include the viscosity of a diffuse interface and stabilize the interface profile in the Stokes-Cahn-Hilliard equation by an auxiliary advection velocity, which is constant normal to the interface. 
The method is numerically validated with previous results based on linear stability analysis. Further, we highlight some distinct features of the new method, like the avoidance of re-meshing and the inclusion of contact mechanics, as we simulate the self-organized polarization and migration of a cell through a narrow channel. Finally, we study the formation of a contractile ring on the surface and illustrate the capability of the method to resolve topological transitions by a first simulation of a full cell division.  
}



\maketitle

\section{Introduction}

The morphogenesis of cells and tissues involves an interplay between chemical signals and the mechanics of their surrounding surface layers \cite{Mayer2010, Salbreux2012, Heisenberg2013, Martin2010}. 
A striking example is found in cell division, during which the cell manages to constrict itself along a ring to finally create two daughter cells \cite{Mayer2010, Pollard2010}. The formation of the contractile ring can be explained by a mechano-chemical mechanism of pattern formation \cite{Mietke_PNAS_2019, Mietke_PRL_2019, Bonati_2022, AlandWittwer_2023}. Thereby, the concentration of contractile surface molecules, such as myosin, spontaneously assembles into a ring shape, driven by the interplay of flow, transport and tension along the surface. 

While various theoretical studies consider self-organized active fluids on fixed domains \cite{Bois2011-ev, Kumar2014-gw, Moore2014-dm, Sehring2015-hg, Weber2018-nn, Mietke_PRL_2019}, the understanding of the pattern formation and active dynamics of \textit{evolving} surfaces requires advanced numerical simulation techniques.
Deforming active surfaces were considered in \cite{Mietke_PNAS_2019, torres2019modelling, Bonati_2022, AlandWittwer_2023}, coupled to hydrodynamics and material flow. Significant shape deformations including strong constrictions could be reproduced for tubular surfaces \cite{Mietke_PNAS_2019} as well as for ellipsoidal and spherical surfaces \cite{Bonati_2022, AlandWittwer_2023}. However, all previous methods operate with a grid-based representation of the surface. Correspondingly, deformations were limited to some extent and topological transitions, such as observed during cell division, were not feasible. Also, the spatial configuration of the surrounding medium was either neglected \cite{Mietke_PNAS_2019, torres2019modelling}, or limited to a simple homogeneous fluid \cite{Bonati_2022, AlandWittwer_2023}. 
To overcome these limitations, we propose a novel phase-field approach to represent the evolving surface. The implicit phase-field description is not only flexible to deal with complex surrounding geometries, including contact to walls and obstacles, but also enables the simulation of topological changes of the active material. 

Phase-field models provide a flexible tool to capture the dynamics of moving interfaces. An auxiliary field variable $\phi$, the phase-field, is introduced and used to indicate the bulk phases, e.g., $\phi=1$ and $\phi=0$, which can be arbitrary viscous, viscoelastic or elastic materials \cite{mokbel2018_PF_FSI}. The phase-field function varies smoothly between these distinct values across the interface, resulting in a small but finite interface thickness. 
Depending on the application of interest, phase-field methods may offer advantages over other interface-capturing methods. For example, they intrinsically include mass conservation and transport-stabilization. Further, they allow for unconditionally stable inclusion of surface-tension \cite{Aland_2014_time} and fully-discrete energy-stable schemes, see e.g. \cite{ChenAland,Gruen}. 
Additional physical processes can be coupled to the multi-phase system by means of the diffuse-interface approach \cite{li2009_diffuse_domain,  RaetzVoigt_2006}, for example to describe interfacial particles \cite{Aland_2011_colloid} or convection-diffusion systems on the interface \cite{Aland_2016_endocytosis,Garcke2014a}. 

In this paper, we develop the first phase-field model to describe deforming active surfaces. Thereby, we couple the surface concentration of a contractile species to the viscous hydrodynamics of the deforming surface and the surrounding viscous media. We end up with a diffuse-interface model which approximates the sharp-interface equations from \cite{Mietke_PRL_2019, Bonati_2022, AlandWittwer_2023}.
The model is validated for spherical geometries with the linear stability analysis from \cite{Mietke_PRL_2019}. We further explore the non-linear regime with a focus on large deformations. We illustrate the capability of the model to study migration of polarized cells including contact mechanics in confined spaces. Finally, we provide the first simulation of an active surface which undergoes a topological transition as we consider spontaneous ring formation leading to a split-up, resembling cell division.

\section{Mathematical Model}
We start by introducing the sharp-interface model of an active gel surface embedded in a surrounding fluid medium. To this end, we formulate the model described in \cite{Mietke_PRL_2019,Mietke_PNAS_2019,AlandWittwer_2023} in a distribution form which is amenable to be transformed into a diffuse-interface formulation.  
Afterwards, these equations are non-dimensionalized and a phase-field formulation is introduced.

\input{2.1_mathematical_model}
\input{2.2_nondimensionalization}
\input{2.3_phase_field_approach}

\section{Numerical Discretization}
\input{3.1_time_discr}
\input{3.2_space_discr}

\section{Numerical Results}
\input{4.1-1_validation_setup}

\input{4.1-3_convergence_exp}
\input{4.2_swimmer}

\input{4.3_division}

\section{Conclusion}
We have presented a phase-field model to describe pattern formation and shape dynamics of active deformable surfaces. The model couples surface and bulk hydrodynamics to surface flow of a diffusible species, which generates an active contractile force. The corresponding sharp-interface equations are carried over to a diffuse interface described by a phase-field, which evolves according to the Stokes-Cahn-Hilliard equation. To maintain a proper interface profile despite strong surface compression or extension, we introduced an auxiliary advection velocity in the Cahn-Hilliard equation and the concentration equation accordingly. This field extends the interfacial velocity constant in the normal direction and is computed by an elliptic PDE, justified by asymptotic analysis. Another new feature is the inclusion of surface viscosity in the diffuse-interface framework, which is necessary for typical biological active surfaces. For example, the viscosity of the cortical cell surface (in Pa\,s\,m) exceeds the intracellular viscosity (in Pa\,s) by more than the typical cell size. 

The new method was validated by numerical convergence tests. In particular, we showed that the interface thickness $\varepsilon$ is small enough to influence the results only marginally. Also a comparison of the obtained patterns with predictions from linear stability analysis \cite{Mietke_PRL_2019} showed good agreement in the obtained critical P\'eclet numbers. Also we note, that the method maintains stable at time step sizes which are orders of magnitude larger than in \cite{Bonati_2022, AlandWittwer_2023} for the same test problems. This is due to the fact that the phase-field model admits an easy monolithic coupling of surface tension, surface viscosity and surface advection.  

Finally, we highlighted some distinct advantages of the new model as compared to previous numerical methods to describe active surface dynamics. We demonstrated the migration of a polarized cell through a fluidic channel, which is possible here without re-meshing and without any effort to realize contact mechanics. 
At last, we considered the division of a cell due to formation of a contractile ring around its periphery. We illustrated that the method is capable to simulate the complete split-up of the cell into two daughter cells, which provides a first simulation of a topological transition of an active surface material. 

The developed method provides a basis to analyze a variety of systems that involve mechanochemical
pattern formation on active surfaces in different surroundings. For example, a detailed study of cell migration in various geometries including obstacles would be worthwhile to understand the general migration modes of cells propelled by rear contraction. 
Also the presence of multiple cells can be described by multiple phase-fields (see \cite{marth2016margination}) to analyze their mechanical interaction during division.
Finally let us note that, while we have described the surface as a viscous material, the constitutive relations can be readily adapted, following \cite{eloy2021numerical}, to account for the viscoelastic behavior which cells show on shorter time scales, in the future. 

\vspace{0.5cm}
\textbf{Acknowledgements:}
We thank Marcel Mokbel, Simon Praetorius and Lucas Wittwer for support of the project and fruitful discussions. SA acknowledges financial support from the DFG in the context of the Forschergruppe FOR3013, project AL1705/6-2 and project AL1705/3-2. Simulations were performed at the Center for Information Services and High Performance Computing (ZIH) at TU Dresden.

\clearpage

\numberwithin{equation}{section}
\appendix
\section{Appendix}
\input{6.2_extending_v}
\input{6.3_asymptotc_analysis}
\input{6.1_axisymmetry}

\bibliography{main}

\end{document}

%% file: 2.1_mathematical_model.tex
\subsection{Sharp-interface model}

The spatial domain, denoted by \(\Omega\), is divided into the intracellular fluid domain \(\Omega_1\) and the exterior fluid medium \(\Omega_0\). Both domains are separated by the cell surface \(\Gamma\) representing the cell membrane and cortex. 
\begin{figure}
    \centering
    \includegraphics[width=10cm]{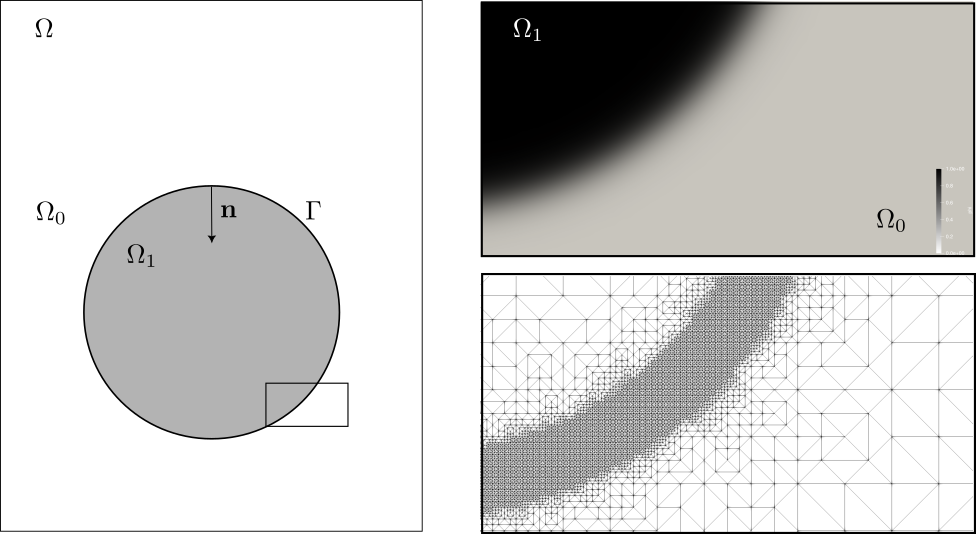}
    \caption{The simulation domain. {\bf Left}: Sketch of the sharp-interface representation, {\bf Right}: Diffuse-interface representation of the inset region by a phase-field $\phi$ (top) and the corresponding adaptive mesh (bottom)} 
    \label{fig:sketch}
\end{figure}

We refer to the normal vector pointing from \(\Omega_0\) into \(\Omega_1\) by \(\mathbf{n}\), and use it to define the projection \(P_\Gamma\) onto \(\Gamma\) by
\begin{align*}
    P_\Gamma \colon = I-\mathbf{n}\otimes\mathbf{n},
\end{align*}
with \(I\) denoting the identity matrix.
This allows us to properly define differential operators on the surface \(\Gamma\) as follows
\begin{align*}
    &\nabla_\Gamma \coloneqq P_\Gamma \nabla &&\text{surface gradient},\\
    &\nabla_\Gamma \cdot \coloneqq P_\Gamma : \nabla &&\text{surface divergence},\\
    &\Delta_\Gamma \coloneqq \nabla_\Gamma \cdot \nabla_\Gamma && \text{Laplace-Beltrami operator}.
\end{align*}
Note, that applying these operators to a field variable requires the latter to be defined not only on $\Gamma$ but in a neighborhood of $\Gamma$. Throughout this article, all surface variables are assumed to be defined in this way.

At the small scales of biological cells ($\upmu$m), gravitational and inertial forces are negligible. Therefore, the  Stokes equations govern the flow in \(\Omega\)
\begin{align*} 
    \nabla \cdot \left[\eta(\nabla \mathbf{u} + \nabla \mathbf{u}^T )\right]- \nabla p  &= \mathbf{f}_\sigma + \mathbf{f}_\text{visc} , \\
    \nabla\cdot\mathbf{u}&= 0, 
 \end{align*}
where $\eta(x)$ denotes the fluid viscosity, $\eta(x)= \eta_i$ for $x\in \Omega_i, i=0,1$. Moreover, $\mathbf{f}_\sigma$ and $\mathbf{f}_\text{visc}$ denote the surface tension and surface viscous forces, respectively, to be specified later. It is important to emphasize that we consider the {Stokes} equation in the distributional sense, since we allow the pressure \(p\) to have a discontinuity on the surface \(\Gamma\).

Those equations are coupled to an advection-diffusion equation to describe the concentration \(c\) of the force-generating molecules (e.g. myosin motor proteins), on \(\Gamma\),
\begin{align*}
    \partial_t^\bullet c + c \nabla_\Gamma \cdot  \mathbf{u} - D \Delta_\Gamma c = 0,
\end{align*}
where $ \partial_t^\bullet$ is the material derivative and $D>0$ the diffusion constant.

It remains to specify the surface forces. The surface tension force $\mathbf{f}_{\sigma}$ is induced by the contraction of force-generating molecules. Similar as in previous literature \cite{AlandWittwer_2023,Mietke_PNAS_2019, Mietke_PRL_2019}, we model the tension in terms of a monotonically increasing {Hill}-function \(\sigma(c) =\xi(\gamma + \frac{c^2-c_0^2}{c^2+c_0^2})\), where \(c_0\) is the  characteristic (equilibrium) concentration, \(\gamma\in\mathbb{R}\) a constant equilibrium surface tension, and $\xi>0$ a scaling factor. Consequently, the surface tension force can be formulated using a surface Dirac-delta function $\delta_\Gamma$ as
\begin{align}
    {\bf f}_\sigma &= \delta_\Gamma H \sigma(c) {\bf n} + \delta_\Gamma \nabla_\Gamma \sigma(c),
\end{align}
 where \(H= \nabla \cdot \mathbf{n}\) is the total curvature.
 Note, that the last term (also called Marangoni term) provides a force toward regions of high concentrations which is the main driving force of the system.

Furthermore, it has been shown that the cell cortex is the main contributor to the mechanical integrity of the cell. Hence, any flow and shape deformation along the surface is limited by the cortical ability to deform and remodel itself, such that accounting for the cortex mechanics is indispensable. On the relevant time scales of cell division and cell migration, the rheology of the cortex can be assumed to be viscous \cite{Bonati_2022}, which gives rise to the surface viscous force \({\bf f}_{\text{visc}}=\nabla_\Gamma  \cdot S_{\text{visc}}\). The stress $S_{\text{visc}}$ of  a viscous surface was introduced by Scriven  \cite{scriven1960dynamics} as
\begin{align*}
    S_{\text{visc}}= (\eta_b-\eta_s)(\nabla_\Gamma \cdot \mathbf{u})P_\Gamma +\eta_s P_\Gamma(\nabla_\Gamma \mathbf{u} + \nabla_\Gamma \mathbf{u}^T)P_\Gamma,
\end{align*} 
where \(\eta_b\) and \(\eta_s\) are the bulk and shear viscosity of the surface.

Also, the surface viscous force must be restricted to the interface. This can be realized by multiplying a Dirac-delta function with either ${\bf f}_\text{visc}$ or ${S}_\text{visc}$. Both approaches are actually formally equivalent as can be seen when regarding the weak formulation for any test function \(\varphi \in H_0^1(\Omega)^3\):
\begin{align*}
  \int_\Omega \delta_\Gamma {\bf f}_\text{visc} \cdot \varphi \dd x 
  &= \int_\Gamma {\bf f}_{\text{visc}} \cdot \varphi \dd x 
  = \int_\Gamma \varphi \cdot \nabla_\Gamma \cdot S_{\text{visc}} \dd x
  = -\int_\Gamma  \nabla_\Gamma \varphi : S_{\text{visc}} \dd x 
  = -\int_\Omega  \delta_\Gamma \nabla_\Gamma \varphi : S_{\text{visc}} \dd x\\
  &= \int_\Omega \nabla_\Gamma \cdot (\delta_\Gamma S_{\text{visc}}) \cdot \varphi  \dd x .
\end{align*}
Here we choose the latter approach and obtain the system: 
\begin{align}
    -\nabla \cdot \left[\eta(\nabla \mathbf{u} + \nabla \mathbf{u}^T )\right] + \nabla p =&
    \nabla_\Gamma \cdot \left[\delta_\Gamma(\eta_b-\eta_s)P_\Gamma\nabla_\Gamma \cdot \mathbf{u} +\eta_s \delta_\Gamma P_\Gamma(\nabla_\Gamma \mathbf{u} + \nabla_\Gamma \mathbf{u}^T)P_\Gamma  \right] \nonumber\\
    &+\delta_\Gamma H \sigma(c) {\bf n} + \delta_\Gamma \nabla_\Gamma \sigma(c) &\text{on }\Omega, \label{eq:sharp_interface:u_stress}\\
    \nabla\cdot\mathbf{u}=& 0&\text{on }\Omega,\label{eq:sharp_interface:u_div}\\
    \partial_t^\bullet c + c \nabla_\Gamma \cdot \mathbf{u} - D \Delta_\Gamma c =& 0 &\text{on }\Gamma. \label{eq:sharp_interface:conc}
\end{align}

%% file: 2.2_nondimensionalization.tex
\subsection{Non-dimensionalization}
To reduce the number of model parameters, the governing equations are non-dimensionalized. To do so, we replace the length scales by \(x^*=\frac{x}{R}\) where \(R\) is the initial radius of the cell and \(t^*=\frac{t}{\tau_D}\) with \(\tau_D=\frac{R^2}{D}\), \(c^*=\frac{c}{c_0}\) with the average concentration \(c_0\) on the surface and \(\mathbf{u^*}=\frac{R}{D}\mathbf{u}\). Additionally, we use the dimensionless version of the differential operators and obtain for the concentration equation
\begin{align*}
    \frac{d(c_0 c^*)}{d(\tau_D t^*)} - D \frac{\Delta_\Gamma^*}{R^2} (c_0c^*)+\frac{\nabla_\Gamma^*}{R} \cdot \left(\frac{c_0 D}{R} c^*\mathbf{u^*}\right)=0.
\end{align*}
Dividing by \(\frac{c_0R^2}{D}\) and omitting the * yields the dimensionless concentration equation
\begin{align}
    \partial_t c + \nabla_\Gamma \cdot (c \mathbf{u})- \Delta_\Gamma c &= 0 &&\text{on }\Gamma. \label{eq:c nondim}
\end{align}
Now, doing the same for the {Stokes}-equation and using $\delta_\Gamma=\delta_\Gamma^*/R$, we obtain
\begin{align*}
    -\frac{\nabla^*}{R} \cdot &\eta \left[ \frac{\nabla^*}{R}\frac{D}{R}\mathbf{u^*}+\left(\frac{\nabla^*}{R}\frac{D}{R}\mathbf{u^*}\right)^T \right]
    +\frac{\nabla^*}{R} p = \frac{\delta_\Gamma^*}{R} \frac{H^*}{R} \sigma(c) {\bf n} + \frac{\delta_\Gamma^*}{R} \frac{\nabla^*_\Gamma}{R} \sigma(c) \\
    &+\frac{\nabla_\Gamma^*}{R} \cdot \left[ (\eta_b-\eta_s) \frac{\delta_\Gamma}{R}P_\Gamma \frac{\nabla_\Gamma^*}{R} \cdot \frac{D}{R}\mathbf{u^*}  +\eta_s \frac{\delta_\Gamma}{R} P_\Gamma\left( \frac{\nabla_\Gamma^*}{R}  \frac{D}{R}\mathbf{u^*}+\left(\frac{\nabla_\Gamma^*}{R} \frac{D}{R}\mathbf{u^*}\right)^T\right) P_\Gamma\right].
\end{align*}
Choosing now \(p^*=\frac{R^3}{D\eta_b}p, \eta^* =\frac{\eta R}{\eta_b}, \gamma^*=\gamma/\xi, \nu = \frac{\eta_s}{\eta_b}\) and introducing the {P\'eclet} number \(\Pe = \frac{\xi R^2}{D\eta_b}\), dividing by \(\frac{D \eta_b}{R^4}\) and again omitting the * gives
\begin{align}
    -\nabla \cdot \left[\eta(\nabla \mathbf{u} + \nabla \mathbf{u}^T )\right] + \nabla p =&
    \nabla_\Gamma \cdot \left[\delta_\Gamma(1-\nu)P_\Gamma\nabla_\Gamma \cdot \mathbf{u} +\nu \delta_\Gamma P_\Gamma(\nabla_\Gamma \mathbf{u} + \nabla_\Gamma \mathbf{u}^T)P_\Gamma  \right] \nonumber\\
    &+\Pe\delta_\Gamma \left(H \sigma(c) {\bf n} + \nabla_\Gamma \sigma(c)\right) &\text{on }\Omega, \label{eq:NS 1 nondim} \\
    \nabla\cdot\mathbf{u}=& 0&\text{on }\Omega, \label{eq:NS 2 nondim}
\end{align}
with 
\begin{align}
    \sigma(c) = \gamma + \frac{c^2-1}{c^2+1}.
\end{align}
The system contains the four parameters $\Pe, \gamma, \nu$ and $\eta$. Note that the latter is the non-dimensional fluid viscosity given by $\eta_i R/\eta_b$ in $\Omega_i, i=0,1$. 

%% file: 2.3_phase_field_approach.tex
\subsection{Phase-field ansatz}
To flexibly account for large deformations, topological transitions and (wall) contact, we introduce a phase-field version of the above equations. The geometry of the cell is described by a phase-field \(\phi\) such that \(\phi(x)\approx 0\) in $\Omega_0$ and  \(\phi(x)\approx 1\)  in $\Omega_1$ with a smooth transition between the two phases.

The phase-field is initialized by $\phi=0.5(1+\text{tanh}\left(r/(\sqrt{2}\varepsilon) \right)$, where $r$ is the signed distance to $\Gamma$, positive in $\Omega_1$. After initialization the phase-field must be advected with the fluid flow to track changes in the surface geometry. This is achieved by the convective Cahn-Hilliard equation
\begin{align}
    \partial_t \phi - \nabla \cdot (M \nabla \mu) + \nabla \phi \cdot \mathbf{u} & =0 && \text{on } \Omega \label{CH-1}\\
    \mu + \varepsilon \Delta \phi - \varepsilon^{-1}W'(\phi) &=0 && \text{on } \Omega \label{CH-2},
\end{align}
where \(\mu\) denotes the chemical potential and \( W(\phi)=\phi^2(1-\phi)^2\) a double-well potential. The parameter \(\varepsilon>0\) describes the thickness of the interface region, and the mobility \(M>0\) regulates the conservation of a smooth interface profile.


The phase-field representation can not only be used to describe the surface \( \Gamma = \{x | \phi(x)=\frac{1}{2} \} \), but also to  {approximate} the Dirac-delta function \(\delta_\Gamma \approx \vert \nabla \phi \vert\).
Following the diffuse-interface approach \cite{RaetzVoigt_2006}, the concentration equation \eqref{eq:c nondim} can be extended from the submanifold \(\Gamma\) onto \(\Omega\) as 
\begin{align}
     \partial_t (\vert \nabla \phi\vert c) + \nabla\cdot (\vert \nabla \phi\vert c \mathbf{u})- \nabla \cdot (\vert \nabla \phi\vert \nabla c) &=0 &\text{ on } \Omega. \label{eq:phase_field:conc}
\end{align}
See Appendix \ref{sec:diffuse interface derivation} for a derivation.
Even though the concentration equation suggests mass conservation, numerical discretization errors may lead to small errors, which can accumulate to a significant mass loss over long simulation times. 
Therefore, we introduce the mass on the surface 
\[m(t) = \int_\Omega \vert \nabla\phi(t)\vert c(t) \dd x \approx \int_{\Gamma(t)} c(t) \dd \sigma\] 
and add  a \textit{mass correction term}
\[\frac{\delta_m}{\Delta t} \vert \nabla \phi \vert (m(t)-m(0))\]
with some \(\delta_m>0\) to the right hand side of the concentration equation to secure mass conservation.

Finally, we reformulate the momentum equation \eqref{eq:NS 1 nondim} in the phase-field formalism. The fluid viscosity is linearly interpolated between the distinct viscosities in the phases, \(\eta(\phi)=(1-\phi)\eta_0 R/\eta_b+\phi\eta_1 R/\eta_b\).
As usual for diffuse-interface models of two-phase flow (e.g. \cite{Aland_2014}), the constant surface tension term $\delta_\Gamma H\sigma {\bf n}$ can be reformulated to $3\sqrt{2}\sigma\mu\nabla\phi$, where the scaling factor emerges due to the chosen double-well potential.
Moreover, we use the extended normal vector \( \mathbf{\tilde n} \coloneqq {\nabla \phi}/{\vert \nabla \phi \vert} \approx \mathbf{n}\) and the surface projection \( \tilde P_\Gamma \coloneqq I- \mathbf{\tilde n} \otimes \mathbf{\tilde n} \approx P_\Gamma \) to define diffuse-interface {approximations} to the surface differential operators, e.g. the surface gradient ${\tilde \nabla}_\Gamma \coloneqq \tilde P_\Gamma \nabla$ and surface divergence 
\({\tilde \nabla}_\Gamma \cdot \coloneqq \tilde P_\Gamma :\nabla \).
We obtain a diffuse-interface version of Eq.~\eqref{eq:NS 1 nondim}
\begin{align}
    -\nabla \cdot \left[\eta(\nabla \mathbf{u} + \nabla \mathbf{u}^T )\right] + \nabla p =&
    {\nabla} \cdot \left[|\nabla\phi|(1-\nu){\tilde P}_\Gamma{\tilde \nabla}_\Gamma \cdot \mathbf{u} + |\nabla\phi| \nu{\tilde P}_\Gamma({\nabla} \mathbf{u} + {\nabla} \mathbf{u}^T){\tilde P}_\Gamma  \right] \nonumber\\
    &+\Pe\left(3\sqrt{2}\sigma(c)\mu\nabla\phi + |\nabla\phi| \sigma'(c){\tilde \nabla}_\Gamma c\right) &\text{in }\Omega,\label{eq:NS 1 phase-field} 
\end{align}
Note that the first divergence operator on the right hand side does not require to be a surface divergence, since it is applied to a tangential tensor.


As we will see in the numerical tests, the active surface tension imposes strong tangential flows leading to regions of large tangential compression ($\nabla_\Gamma \cdot {\bf u}<0$) or stretching ($\nabla_\Gamma \cdot {\bf u}>0$). 
Due to incompressibility of the flow field, this goes along with the opposite deformation in the normal direction, i.e. tangentially stretched regions get compressed in the normal direction and vice versa. 
In numerical tests we find that such strong compressional or extensional flows in the normal direction, tend to locally shrink or widen the thickness of the interface region, respectively, unless an extremely high, unphysical mobility is used.  
To eliminate this perturbation of the interface profile for reasonable mobilities, we instead advect the phase-field with an auxiliary velocity field ${\bf v}$.
The idea is that ${\bf v}$ is an extension of the velocity at the surface (i.e. at the 0.5-level set), which is constant in the normal direction such that the whole interfacial area is advected with the surface velocity. We propose to achieve this extension by solving the additional equation
\begin{align}
    \vert \nabla \phi \vert \mathbf{v}-\vert \nabla \phi \vert\mathbf{u}-\nabla \cdot [\vert \nabla \phi \vert( \mathbf{\tilde n} \times \mathbf{\tilde n}) \nabla \mathbf{v}] &= \mathbf{0} &\text{on }\Omega. \label{eq:phase_field:v_extended}
\end{align}
In the Appendix \ref{sec:asymptotic} we show by matched asymptotic expansion that this formulation converges to the following sharp-interface-limit equations 
\begin{align}
    \mathbf{v} &= \mathbf{u}&\text{on }\Gamma,\label{eq:sharp_interface:u=v}\\
    \nabla \mathbf{v} \cdot \mathbf{n} &= 0 &\text{on }\Gamma.\label{eq:sharp_interphase:v'=0}
\end{align}
The obtained velocity field \(\mathbf{v}\) is used to replace the velocity \(\mathbf{u}\)  in the advection terms of both - the Cahn-Hilliard equation and the concentration equation.  
Consequently, we obtain the final coupled system to be solved throughout \(\Omega\):
\begin{align*}
    \partial_t \phi =& \nabla \cdot (M \nabla \mu) - \nabla \phi \cdot \mathbf{v},\\
    \mu =& -\varepsilon \Delta \phi + \varepsilon^{-1}W'(\phi) ,\\
    -\nabla \cdot \left[\eta(\phi)(\nabla \mathbf{u} + \nabla \mathbf{u}^T )\right] + \nabla p =&
    \nabla \cdot \left[|\nabla\phi|(1-\nu){\tilde P}_\Gamma{\tilde \nabla}_\Gamma \cdot \mathbf{u} +\nu |\nabla\phi| {\tilde P}_\Gamma(\nabla\mathbf{u} + \nabla\mathbf{u}^T){\tilde P}_\Gamma  \right] \nonumber\\
    &+\Pe\left(3\sqrt{2}\sigma(c)\mu\nabla\phi + |\nabla\phi| \sigma'(c){\tilde \nabla}_\Gamma c\right) ,\\
    \nabla\cdot\mathbf{u}=& 0,\\
    \vert \nabla \phi \vert \mathbf{v}=& \vert \nabla \phi \vert\mathbf{u}+\nabla \cdot [\vert \nabla \phi \vert(\mathbf{\tilde n} \times \mathbf{\tilde n}) \nabla \mathbf{v}] ,\\
    \partial_t (\vert \nabla \phi\vert c ) =& -\nabla \cdot (\vert \nabla \phi\vert c \mathbf{v})+ \nabla \cdot (\vert \nabla \phi\vert \nabla c) +\delta_m \vert \nabla \phi \vert (m(t)-m(0)) ,
\end{align*}
where \(\mathbf{\tilde n} = \frac{\nabla \phi}{\vert \nabla \phi \vert} \) and \(\tilde P_\Gamma = I- \frac{\nabla \phi \otimes \nabla \phi}{\nabla \phi \cdot \nabla \phi}\).

%% file: 3.1_time_discr.tex
\subsection{Time Discretization}
After establishing the equation system it remains to solve the six strongly coupled equations. To reduce the size of the linear equation system that we have to solve in each time step, we decouple them and solve the Stokes-Cahn-Hilliard equations independently from the \(\mathbf{v}\) equation and the concentration equation.

We employ the stable linear semi-implicit time discretization from \cite{Aland_2014} for solving the coupled Stokes-Cahn-Hilliard-Navier system in the n-th time step
\begin{align*}
    \frac{\phi^n-\phi^{n-1}}{\Delta t} \phi =& \nabla \cdot (M \nabla \mu^n) - \nabla \phi^n \cdot \mathbf{v}^{n-1},\\
    \mu^n =& -\varepsilon \Delta \phi^n + \varepsilon^{-1}[W'(\phi^{n-1}) + W''(\phi^{n-1})(\phi^n-\phi^{n-1})] ,\\
    -\nabla \cdot \left[\eta(\phi^{n-1})(\nabla \mathbf{u}^n + (\nabla \mathbf{u}^n)^T )\right] + \nabla p^n =&
    Pe\left(3\sqrt{2}\sigma(c^{n-1})\mu^n\nabla\phi^{n-1} + |\nabla\phi^{n-1}| \sigma'(c^{n-1}){\tilde \nabla}^{n-1}_\Gamma c^{n-1}\right) \\
    + \nabla \cdot \left[|\nabla\phi^{n-1}|(1-\nu){\tilde P}^{n-1}_\Gamma{\tilde \nabla}^{n-1}_\Gamma\right.&\cdot \mathbf{u}^n +\left.\nu |\nabla\phi^{n-1}| {\tilde P}^{n-1}_\Gamma(\nabla \mathbf{u}^n + \nabla (\mathbf{u}^n)^T){\tilde P}^{n-1}_\Gamma  \right],\\
    \nabla\cdot\mathbf{u}^n=& 0. 
\end{align*}
where we calculated \(\mathbf{v}^{n-1}\) and \(c^{n-1}\) in the previous time step. 

Using now the current time steps \(\phi^n\) and \(\mathbf{u}^n\), we obtain the projected velocity \(\mathbf{v}\) from
\begin{align*}
    \vert \nabla \phi^{n} \vert \mathbf{v}^n -\nabla \cdot \left(\vert \nabla \phi^{n} \vert({ \mathbf{\tilde n}}^n \times { \mathbf{\tilde n}}^n) \nabla \mathbf{v}^n \right) = \vert \nabla \phi^{n} \vert\mathbf{u}^{n}.
\end{align*}

Finally, we solve the concentration equation using the previously computed values for \(\phi^n\) and \(\mathbf{v}^n\)
\begin{align*}
    \frac{ \vert \nabla \phi^{n}\vert c^n - \vert \nabla \phi^{n-1}\vert c^{n-1}}{\Delta t}+ \nabla \cdot (\vert \nabla \phi^n \vert c^n \mathbf{v}^{n})- \nabla \cdot (\vert \nabla \phi^{n}\vert \nabla c^n) &=\delta_m \vert \nabla \phi^n \vert (m(t^{n-1})-m(0)).
\end{align*}

Note that, even without the term on the right hand side (i.e. in case $\delta_m=0$), the proposed time discretization ensures mass conservation on the discrete level. This property becomes obvious in the weak from of the equation with a test function \(\varphi \in H^1(\Omega)\)
\begin{align*}
      \int_\Omega \varphi \vert \nabla \phi^{n}\vert c^n \dd x- \Delta t \int_\Omega \nabla \varphi \cdot \left[(\vert \nabla \phi^n \vert c^n \mathbf{v}^{n})- (\vert \nabla \phi^{n}\vert \nabla c^n)\right] \dd x = \int_\Omega \varphi \vert \nabla \phi^{n-1}\vert c^{n-1}\dd x.
\end{align*}
This holds especially for \(\varphi\equiv 1\) and thus,
\begin{align*}
   \int_\Omega \vert \nabla \phi^{n}\vert c^n \dd x = \int_\Omega \vert \nabla \phi^{n-1}\vert c^{n-1} \dd x
\end{align*}
which is the phase-field equivalent of exact mass conservation 
\begin{align*}
    \int_{\Gamma^n} c^n \dd x = \int_{\Gamma^{n-1}} c^{n-1} \dd x.
\end{align*}
However, due to adaptive grid refinement and coarsening, small errors in surface mass may accumulate over time, requiring the mass correction ($\delta_m>0$) for long simulations times. 

%% file: 3.2_space_discr.tex
\subsection{Space Discretisation}

We solve the system of equations in each time step with a Finite Element method based on the Finite Element toolboxes DUNE \cite{Sander_2020} and AMDiS \cite{amdis2,VeyVoigt_2006,Witkowski_2015}. To decrease the size of the system, we avoid solving the full 3D-problem and assume that the cell shape and concentration distribution are  axisymmetric. This holds in particular for the biologically most relevant patterns, which are rings and single spots   of increased concentration. The assumption of axisymmetry reduces computations effectively to a 2D domain from which the full 3D-solution can be recovered by rotating the calculated 2D-solution 360 degrees. Detailed explanations can be found in the Appendix \ref{app:axisymmetric}.

An adaptive grid is employed to accurately resolve the phase-field and surface forces. Interfacial grid refinement is heuristically chosen, based on the value of the phase-field, such that the grid size is \( h_{\text{int}}\) where \(0.05<\phi<0.95\) and \( h_{\text{bulk}}\) otherwise. 

The numerical approximations \(\mathbf{\tilde n}\) and \(\tilde P_\Gamma\) become less accurate when evaluated farther from the interface. To avoid numerical errors accumulating in the outer areas, we replace \(\vert \nabla \phi \vert\) by
\[\vert \nabla \phi \vert^* \coloneqq \begin{cases} \vert \nabla \phi \vert, & \text{if } \vert \nabla \phi \vert>10^{-2}\\ 0, &\text{otherwise.}\end{cases}\]
Interchanging \(\vert \nabla \phi \vert^*\) with \(\max(\vert \nabla \phi \vert^*, 10^{-4})\) in the diffusion terms in both equations, we make sure that the induced linear systems remain regular.
For the adjusted problem and the introduced discretization we use Lagrange-P2 elements for the phase-field \(\phi\), the chemical potential \(\mu\), the velocities \(\mathbf{u}\) and \(\mathbf{v}\) and the concentration \(c\), only for the pressure \(p\) we use a P1-ansatz space.

%% file: 4.1-1_validation_setup.tex
\subsection{Numerical Validation}

To validate the derived phase-field model we compare numerical results to the linear stability analysis of \cite{Mietke_PRL_2019}.
Therefore we will analyze the ability of the system to exhibit self-organized formation of concentration patterns, as similarly done in \cite{AlandWittwer_2023}.
The origin of patterns formation is a positive feedback loop between mechanics and molecule transport on the surface: local maxima of myosin concentration on the surface cause contraction towards these regions, which induces a flow transporting more myosin molecules towards the concentration maxima and thus, reinforcing the pattern. 
Eventually, this feedback loop arrives at a steady state, where advection and diffusion are balanced. 

\begin{figure}
    \centering
    \includegraphics[width=0.8\textwidth]{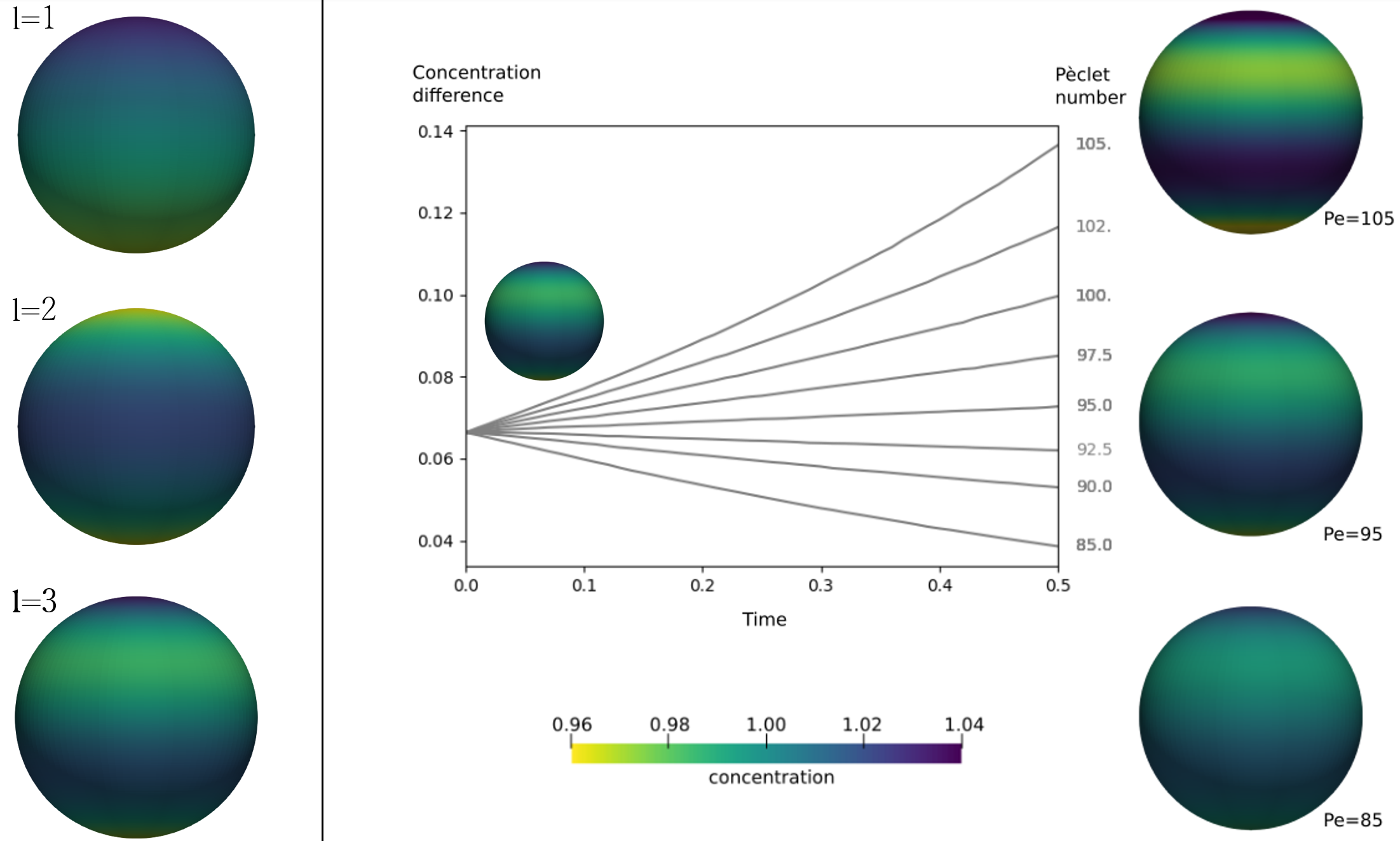}
    \caption{\textbf{Left panel}: Initial concentrations chosen by the first three spherical harmonic modes.
    \textbf{Right panel}:
    Pattern development with initialized $l=3$ mode for \(\eta_0=0.1, \eta_1=10.0\), \(\nu=1\). 
    }
    \label{fig:mode3:evolution}
\end{figure}

The emergence of patterns depends on the ratio between advection and diffusion given by the P\'eclet number \(\Pe\). Pattern formation takes place if this number exceeds a threshold, called the critical P\'eclet number, which also depends on the specific shape of the pattern.  
Simple patterns are given by the first  \textit{eigenmodes} of the linearised differential system and coincide with the \textit{spherical harmonics}, see Fig.~\ref{fig:mode3:evolution}, left panel. 
The \textit{critical} P\'eclet number \(\Pe_l^*\) which is needed for the \(l^\text{th}\) eigenmode to grow has been computed in the linear regime as
\begin{equation}\label{criticalPe2}
    \Pe_l^* =l(l+1)+\nu(l-1)(l+2)+(1+2l){\eta}_\text{bulk},
\end{equation}
see Eq.~(13) in \cite{Mietke_PRL_2019}, in non-dimensional form without attachment. 
The equation has been derived for a viscous cell surface enclosing a viscous interior of (non-dimensional) viscosity $\eta_\text{bulk}$. The presence of the viscous exterior in our model can be effectively taken into account by adding the two viscosities, hence we set $\eta_\text{bulk} = \eta_0+\eta_1$ in the following. 
In our numerical simulations we can observe if patterns grow or decay for various {P\'eclet} numbers, and hence establish a numerical critical {P\'eclet} number to be compared with Eq.~\eqref{criticalPe2}. 
Therfore, we consider a single cell, centered in the axisymmetric domain $[0,3]\times[0,6]$. We compare pattern formation for the modes \(l=1,2,3\) and different combinations of the viscosity parameters \(\eta_\text{bulk} = \eta_0+\eta_1\), \(\nu\) and  {P\'eclet} numbers in an (almost) linear regime. Therefore, we initialize the concentration pattern as the respective target mode with a small amplitude, see Fig. \ref{fig:mode3:evolution}, left panel. 
Using the discretization parameters given in Tab.~\ref{tab:parameters}, we run simulations for 500 time steps and 
quantify the pattern development by the difference between the concentration maximum and minimum on the surface, see Figure \ref{fig:mode3:evolution}. 
\begin{table}[ht]
\centering
\begin{tabular}{lll}
    \hline
    time step size &\(\Delta t\) & \(10^{-3}\)\\    
    grid size &\( h_{\text{bulk}}\) & \(2.65\cdot 10^{-1}\)\\ 
    grid size &\( h_{\text{int}}\) & \(4.14\cdot 10^{-3}\)\\  
    mobility &\(M\) &  \(10^{-3}\)\\
    interface thickness &\(\varepsilon\) & \(0.01\)\\
    mass correction &\(\delta_m\)& 0.25 \\
    equilibrium surface tension & $\gamma$ & 0 \\
    \hline
\end{tabular}
\caption{Parameters used for the validation study.}
\label{tab:parameters}
\end{table}

\begin{figure}
    \centering
    \begin{minipage}{8cm}
    \includegraphics[width=8cm]{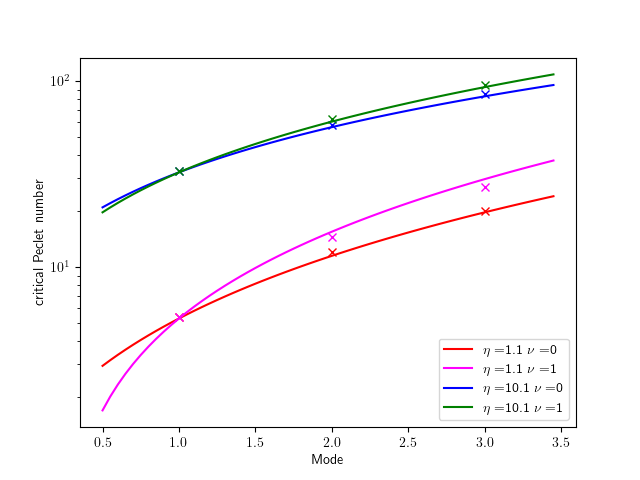}
    \includegraphics[width=8cm]{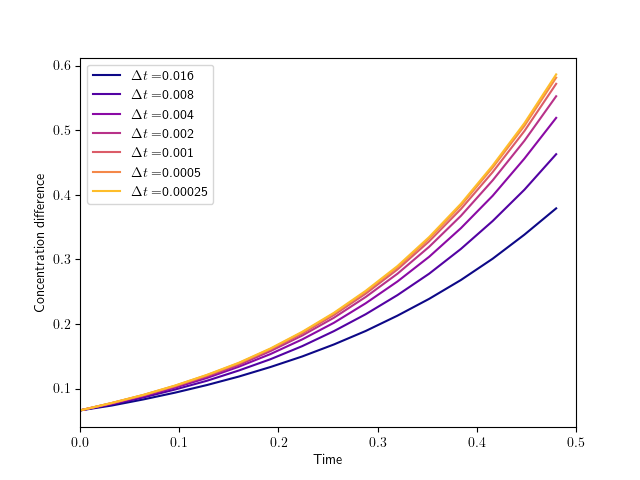}
    \end{minipage}
    \begin{minipage}{8cm}
    \includegraphics[width=8cm]{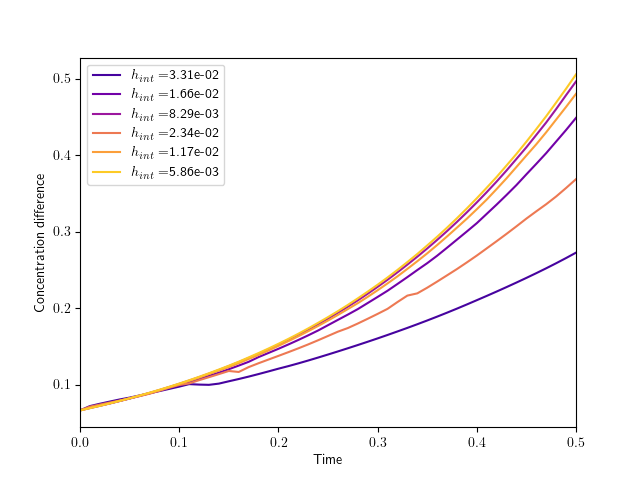}
    \includegraphics[width=8cm]{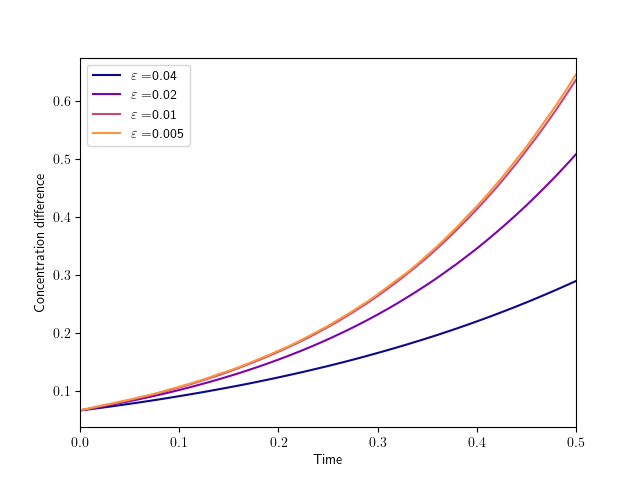}
    \end{minipage}
    \caption{Validation study of the proposed model. \textbf{Top left}: Numerically estimated critical {P\'eclet} numbers (marker points) agree well with the theoretical prediction (lines) from Eq.~\eqref{criticalPe2}. \textbf{Other panels}: Surface concentration difference over time shows convergent behavior in $\Delta t$, interfacial grid size $h_\text{int}$ and interface thickness $\varepsilon$.}
    \label{fig:critical Peclet numbers}
\end{figure}

The numerically estimated critical {P\'eclet} number is obtained as the value at which the pattern remains almost stationary. The resulting values are shown in Fig.~\ref{fig:critical Peclet numbers} (top left). 
We observe that the numerical model reproduces comparable results which agree well with the theoretical prediction from Eq.~\eqref{criticalPe2}. 


%% file: 4.1-3_convergence_exp.tex
Additionally, we perform a study of convergence with respect to the numerical parameters \(\Delta t\), \(h_\text{int}\) and \(\varepsilon\). 
Therefore, we consider the evolution of a \(3\)-mode with \(\eta_0+\eta_1 = 0.1+10\), \(\nu=1\) and \(\Pe=130 > \Pe_3^*\) which results in a growing pattern. 
The evolutions of the concentration difference \(c_{\max} -c_{\min}\) are shown in Fig.~\ref{fig:critical Peclet numbers}. 
Moreover, we compute the experimental order of convergence (EOC) in $\Delta t, h_\text{int}$ and $\varepsilon$, see Tab.~\ref{tab:convergence}. 
As expected, we find convergence of first order in time and second order in space. 
We also find convergence in $\varepsilon$, however the order of convergence is less clear, due to the fixed interface resolution, which might not sufficiently resolve the finest interface thickness. Note, that 
the dependency between grid size and interface thickness to achieve a clear convergence order is a priori not known and part of ongoing research \cite{schmeller2023sharp, Demont2023}. However, we conclude the validation as we obtain convergent behavior with minimal differences between the finest scale solutions. 

\begin{table}[h]
    \centering
    \begin{tabular}{ccc|ccc|ccc}
    \hline
    \(\Delta t\)  & \(c_{\max} -c_{\min}\) & EOC &\(h_\text{int}\)  & \(c_{\max} -c_{\min}\) & EOC & \(\varepsilon \)  & \(c_{\max} -c_{\min}\) & EOC\\
    \hline
    0.008\hspace{5pt}   & 0.46288   &  -    & \(2.34 \cdot 10^{-2}\) &  0.3686\hspace{5pt}  & -     &0.04\hspace{5pt}    & 0.28978 & -\\
    0.004\hspace{5pt}   & 0.51911   &  -    & \(1.17 \cdot 10^{-2}\) &  0.48014 & -     & 0.02\hspace{5pt}    & 0.50869 & -\\
    0.002\hspace{5pt}   & 0.55257   &  0.75 & \(5.86 \cdot 10^{-3}\) &  0.50554 & 2.13  &0.01\hspace{5pt}    & 0.63707 & 0.77\\
    0.001\hspace{5pt}   & 0.5718\hspace{5pt}    &  0.80 & & & & 0.005   & 0.64643 & 3.78\\
    0.0005  & 0.58135   &  1.01 &&&&&\\
   \hline
    \end{tabular}
    \caption{Convergence analysis for refinement of time step size \(\Delta t\), interfacial grid size \(h_\text{int}\) and interface thickness $\varepsilon$. The experimental order of convergence (EOC) is computed from \(c_{\max} -c_{\min}\) at end time \(t=0.5\) for \(h_\text{int}\) and \(\varepsilon\) and \(t=0.48\) for \(\Delta t\). Parameters are as in Tab.~\ref{tab:parameters}, except \(h_\text{bulk}=2.65 \cdot 10^{-1}\), \(h_\text{int}=5.86 \cdot 10^{-3}\) (for \(\Delta t \)-study) and \(\varepsilon=0.02\) (for \(h_\text{int}\)-study).}
    \label{tab:convergence}
\end{table}

%% file: 4.2_swimmer.tex
\subsection{Migration of polarized cells}
As previously predicted in \cite{Mietke_PRL_2019}, the asymmetric patterns of the odd modes produce an asymmetric flow that is able to propel the cell through its surrounding medium, consistent with typical modes of cell crawling \cite{lammermann2009mechanical}. 
Motion is driven in this case by active retrograde flows of the cell surface which stem from contraction of actomyosin at the rear of the cell. 
Numerical simulations of this process have been provided in \cite{AlandWittwer_2023} using a grid-based approach, which is able to represent cell migration in an unobstructed viscous environment. 
However, in the biological setting, cell migration is largely influenced by the geometry of the environment which has a non-trivial influence on the process due to the coupling with flow and pattern formation. 
For example in \cite{Wanda} it was found that confinement can increase the locomotion speed 
during bleb-mediated migration. 

As the proposed phase-field model uses an implicit representation of the cell geometry, it can be easily adapted to simulate cells traversing a complex environment.
We illustrate this in the following by simulating cell migration in free surroundings, as well as the migration into a channel. Both cells have the same mode-1 initial concentration.
Discretization parameters were adapted to account for the longer time interval and the larger spatial movement of the cell:  $\Delta t=10^{-4}, h_{\text{bulk}}=0.25,  h_{\text{int}}=3.12\cdot 10^{-2}, \varepsilon=0.04$.
The channel walls are equipped with no-slip (${\bf v}=0$) and no adhesion ($\phi=0, {\bf n}\cdot\nabla\mu=0$) boundary conditions.


\begin{figure}
    \centering 
    \begin{minipage}{6cm}
    ~~~~ a) Unconstricted cell motion 
    \vspace{-0.9cm}\newline 
    \includegraphics[height=4.9cm, trim={0 3.5cm 0 0},clip]{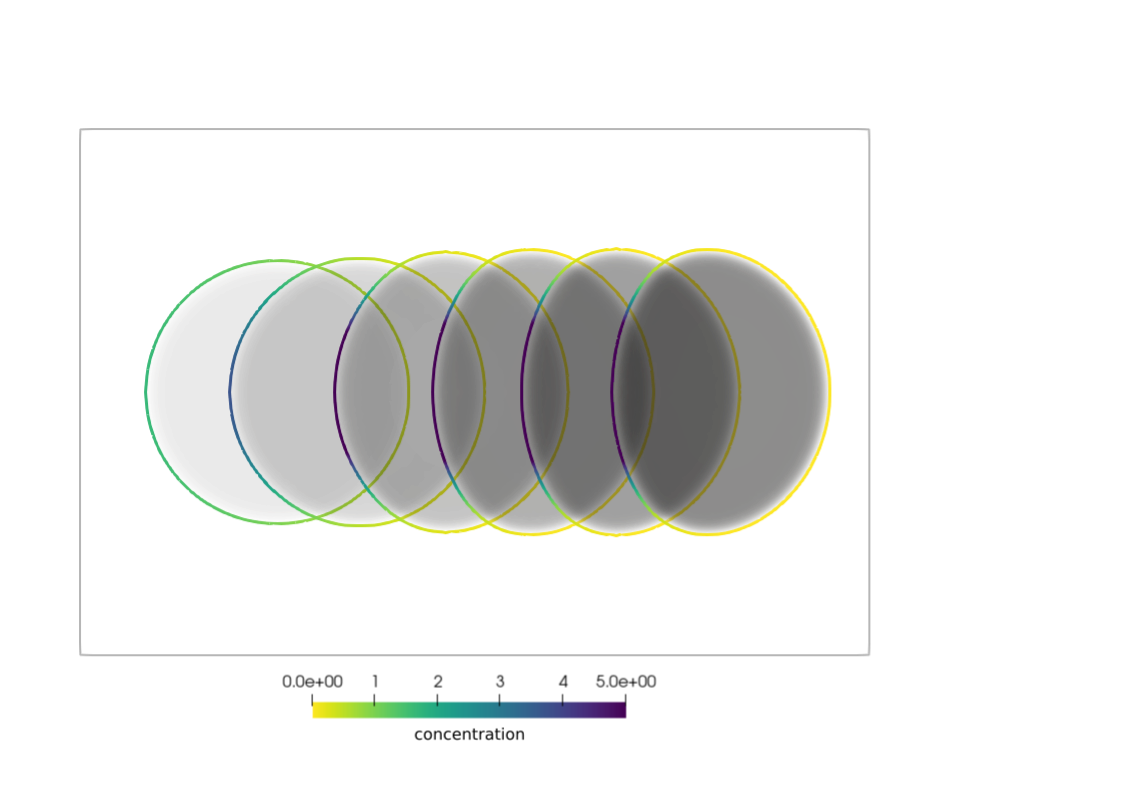}
    \vspace{-1.3cm}\newline
    \includegraphics[height=5.8cm]{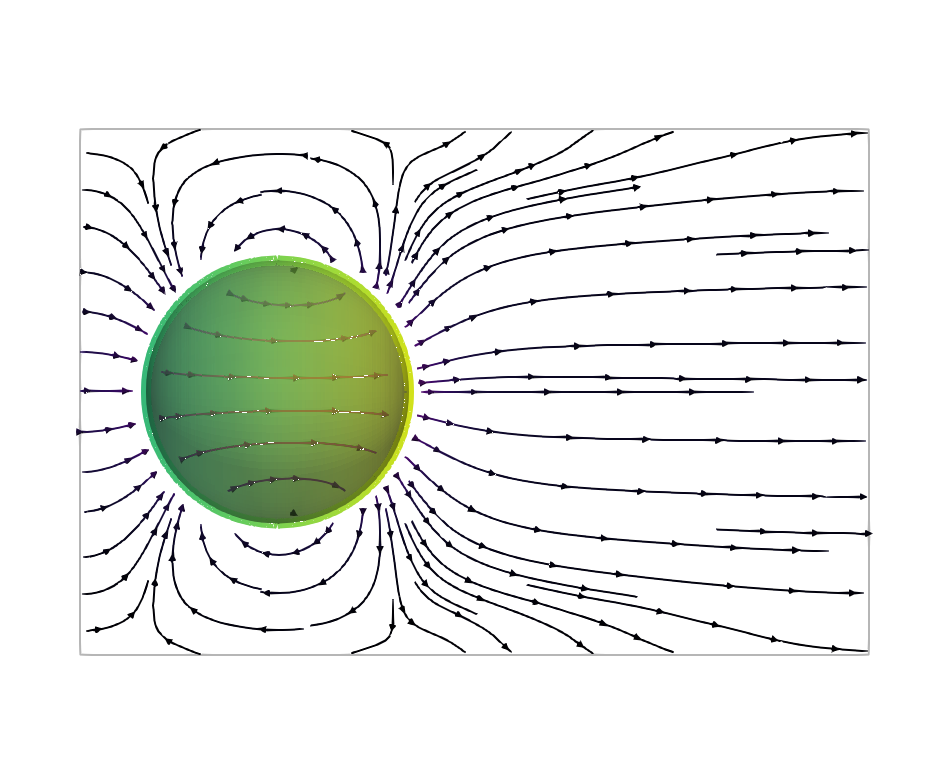}
    \vspace{-2.2cm}\newline
    \includegraphics[height=5.8cm]{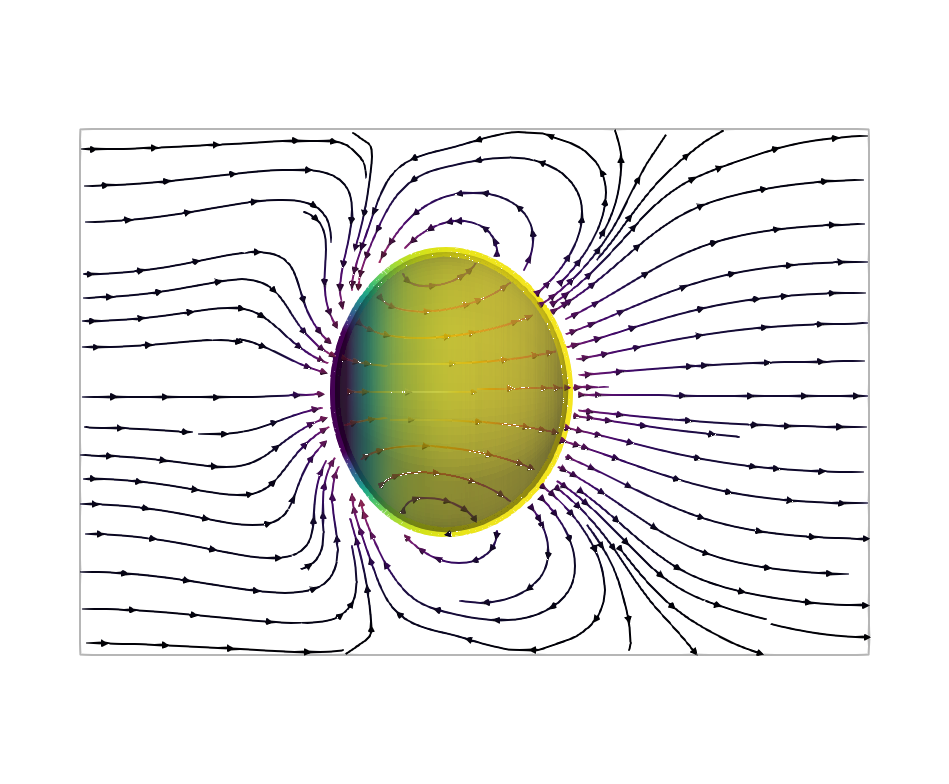}
    \vspace{-2.2cm}\newline
    \includegraphics[height=5.8cm]{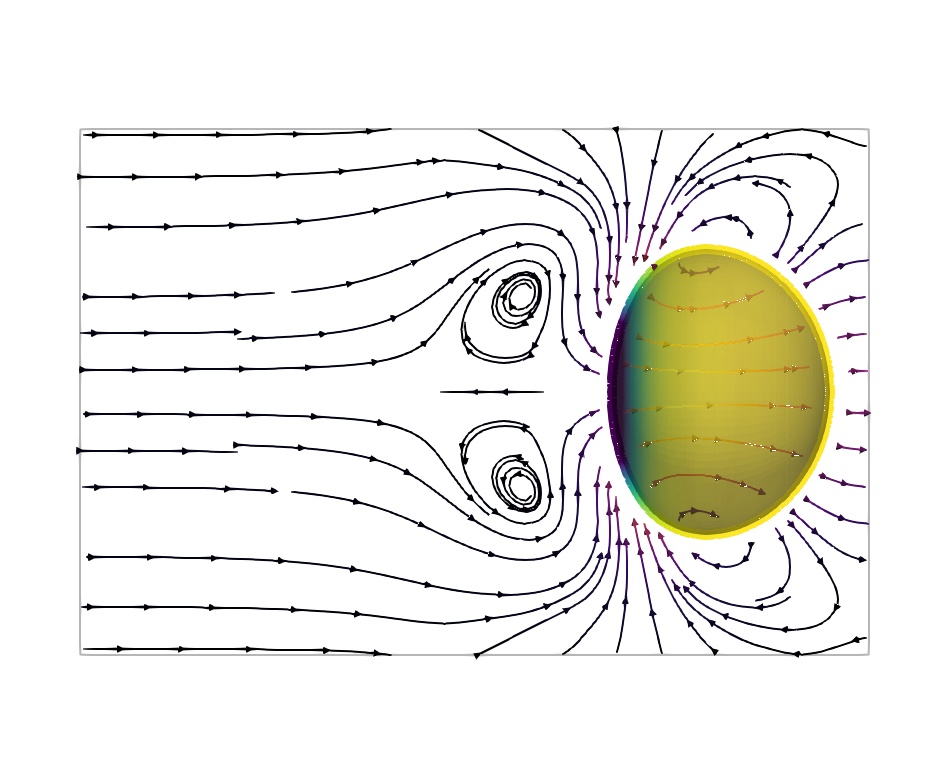}
    \end{minipage}
    \hspace{1cm}
    \begin{minipage}{6cm}
    ~~~~ b) Constricted cell motion 
    \vspace{-0.65cm}\newline 
    \includegraphics[height=4.6cm, trim={0 3.5cm 0 0},clip]{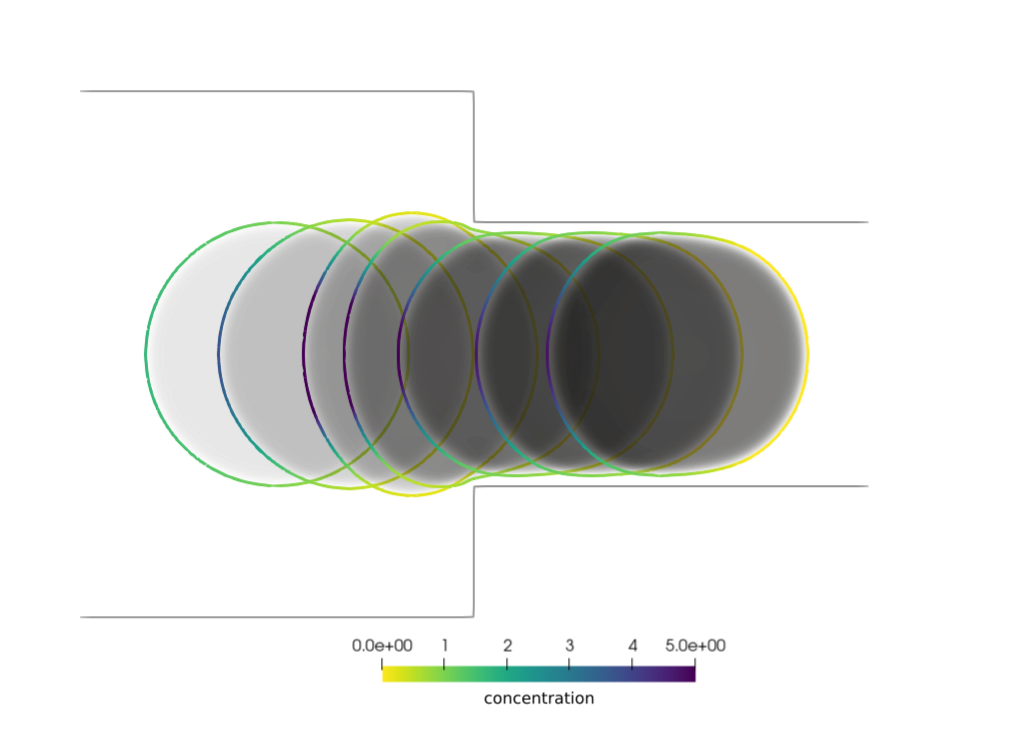}
    \vspace{-1.3cm}\newline 
    \includegraphics[height=5.8cm]{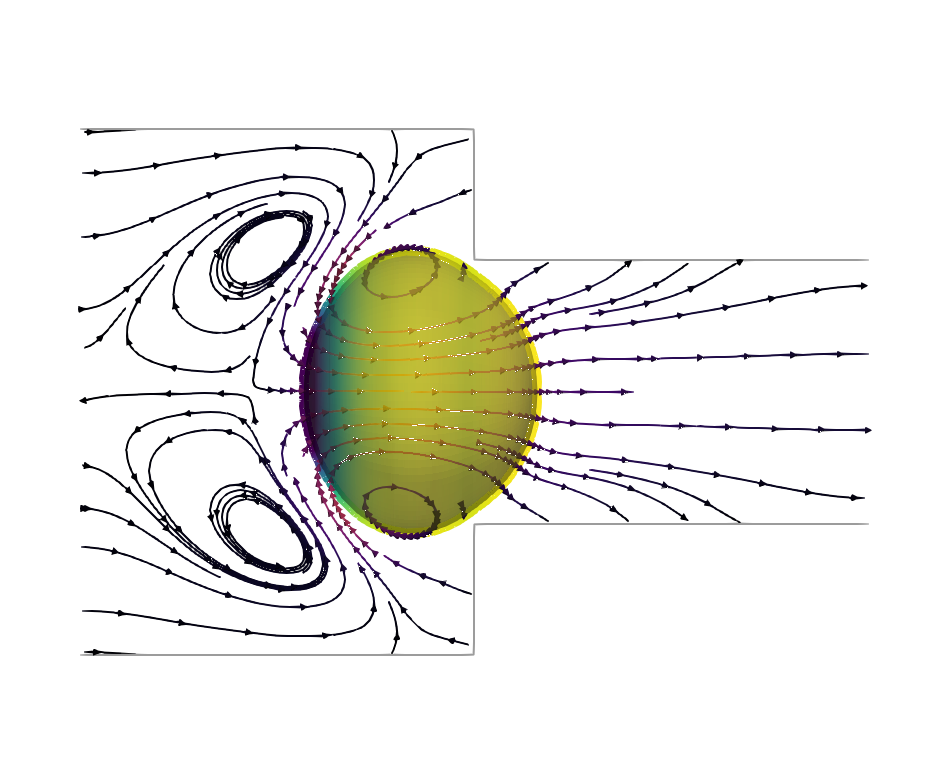}
    \vspace{-2.2cm}\newline
    \includegraphics[height=5.8cm]{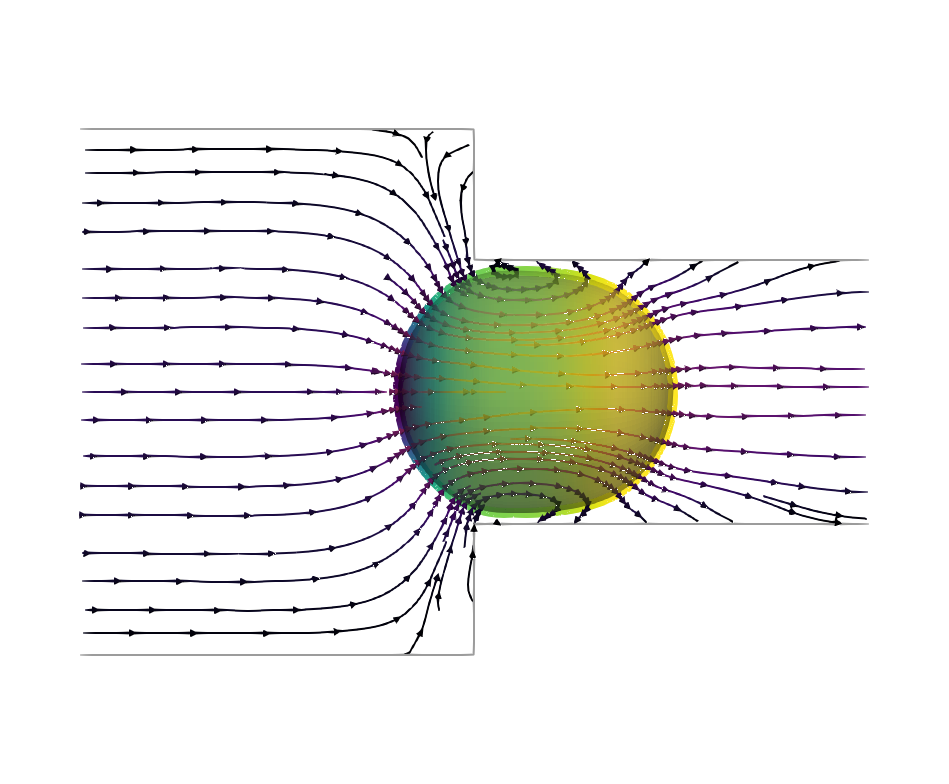}
    \vspace{-2.2cm}\newline
    \includegraphics[height=5.8cm]{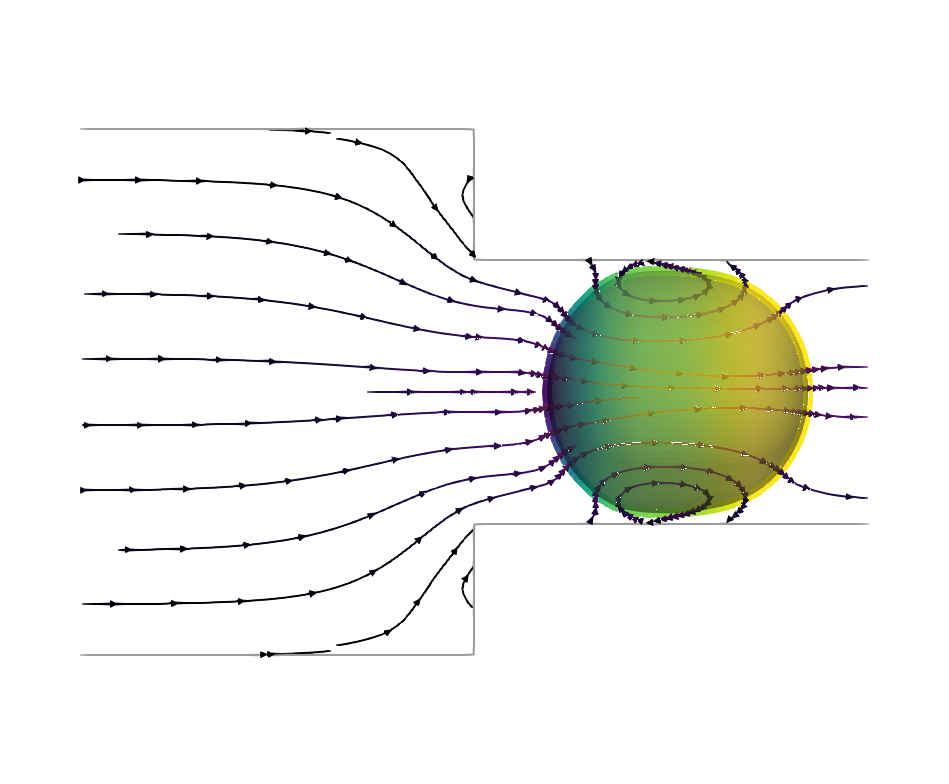}
    \end{minipage}
    \vspace{-1.4cm}\newline
    \includegraphics[width=14cm]{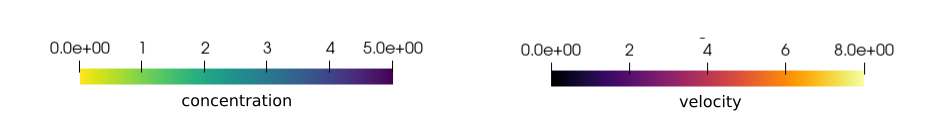}
    \vspace{-0.4cm}
    \caption{Cell migration by rear contraction of actomyosin caused by a polar pattern. {\bf a)} The unconstrained case. Overlay (top) and three snapshots (below) of several times points show formation of a stationary polar pattern. The corresponding contraction at the cell rear propels the cell through the viscous environment at steady speed. Shown time points $t=0.2k, k=0,\ldots, 5$ (top) and $t=0, 0.4, 1.0$ (below). 
    {\bf b)} Overlay (top) and snapshots (below) of several times points of cell migration at the inlet of a constriction. The flow and concentration patterns adapt due to the constriction but lead to ongoing migration of the squeezed cell, with a lower stationary velocity. Shown time points $t=0.2k, k=0,\ldots 6$ (top) and $t=0.4k, k=1,\ldots, 3$ (below).     
    Parameters: $\gamma = 1, \Pe = 20, \nu=1, \eta=1.0+0.1$}
    \label{fig:channel}
\end{figure}

Figure \ref{fig:channel} depicts the pattern formation, cell movement and shape evolution. In the unconstricted case (Fig.~\ref{fig:channel}a), we observe a phase of  acceleration defined by a growing polar pattern and increasing flow magnitude. Meanwhile, the cell transitions into an oblate shape. Eventually, a stationary pattern is reached and the cell migrates at a constant velocity, driven by continuous rear contraction.

In the constricted case (Fig.~\ref{fig:channel}b), the channel size is reduced to a narrow region of half the original diameter, from 4 to 2. As the cell approaches the inlet of the narrow region, the flow and concentration patterns adapt to the geometry. Accordingly, the cell slows down, squashes in the constriction, and migrates further through the narrow passage. The observed stationary pattern is slightly weaker than in the unconstricted case and leads to a lower stationary velocity. 

We note here that the phase-field model is capable of simulating this process without re-meshing and without an explicit handling of the contact mechanics. Consequently, the proposed model permits to study migration in general  environments, including confinement and general channel geometries, in the future.

%% file: 4.3_division.tex
\subsection{Ring slipping and cell division}

Another advantage of the proposed model is the capability to simulate strong deformations and topological transitions. The most radical shape change which a cell undergoes is given by its deformation and fission during the division of a mother cell into two daughter cells.
This process is preceded by the symmetry-breaking formation of a ring of contractile molecules \cite{pollard2010mechanics}. The
theoretical model of Mietke et al. \cite{Mietke_PRL_2019} predicts parameter regions where such a ring may spontaneously emerge in the linear regime, indicated by a dominant $l=2$ mode.

\begin{figure}
    \centering
    \begin{minipage}{7cm}
    ~~~~ a) Ring Slipping 
    \vspace{-1.0cm}\newline 
    \includegraphics[height=6cm]{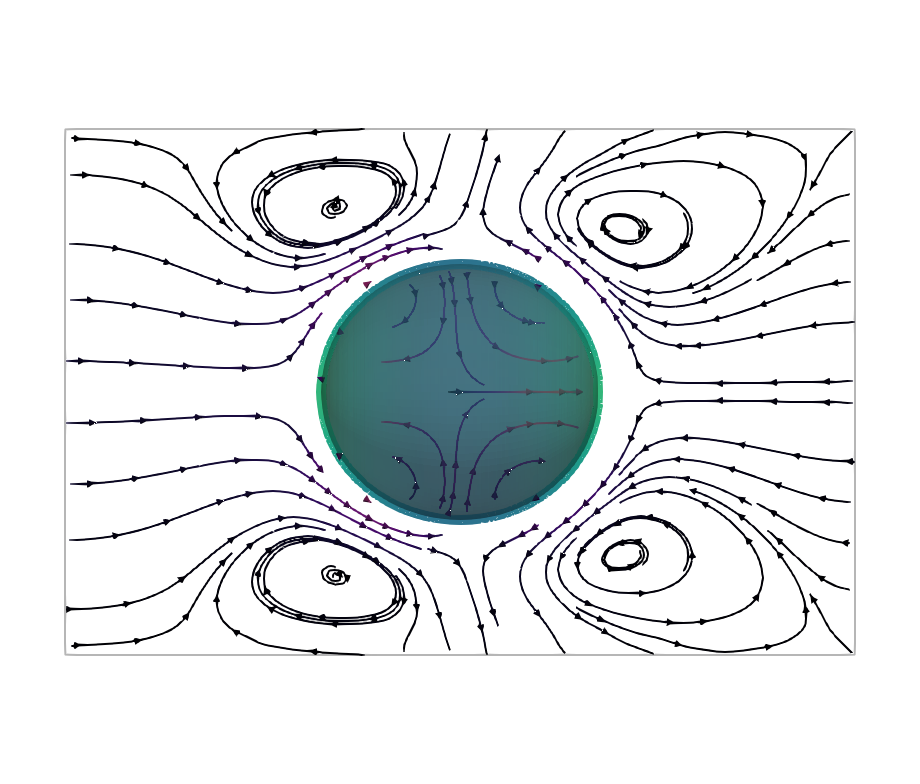}
    \vspace{-2.3cm}\newline 
    \includegraphics[height=6cm]{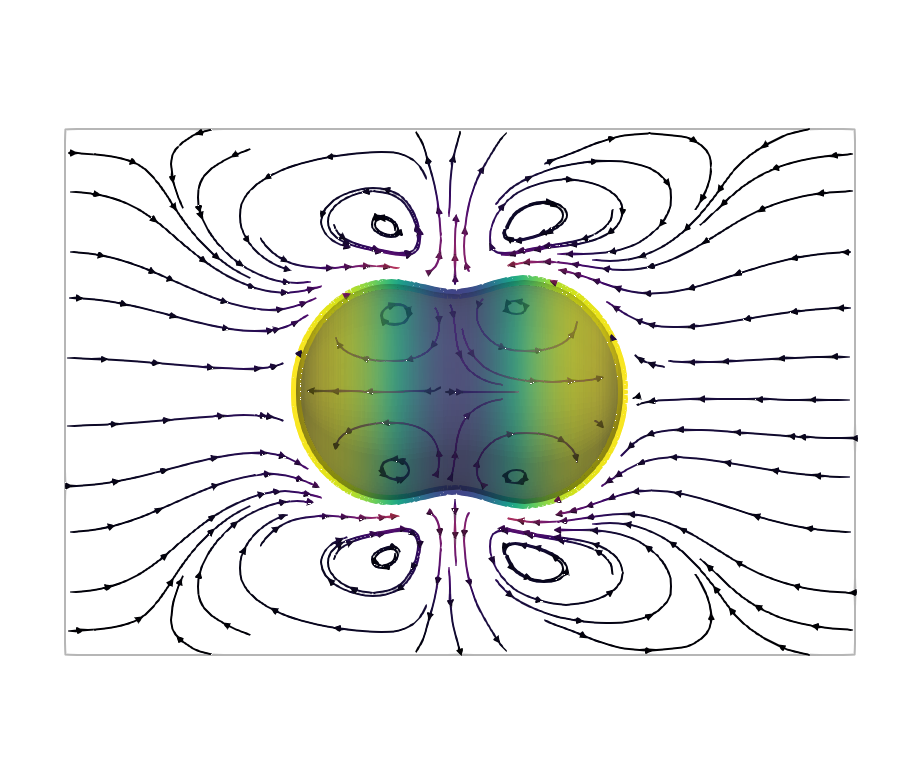}
    \vspace{-2.3cm}\newline 
    \includegraphics[height=6cm]{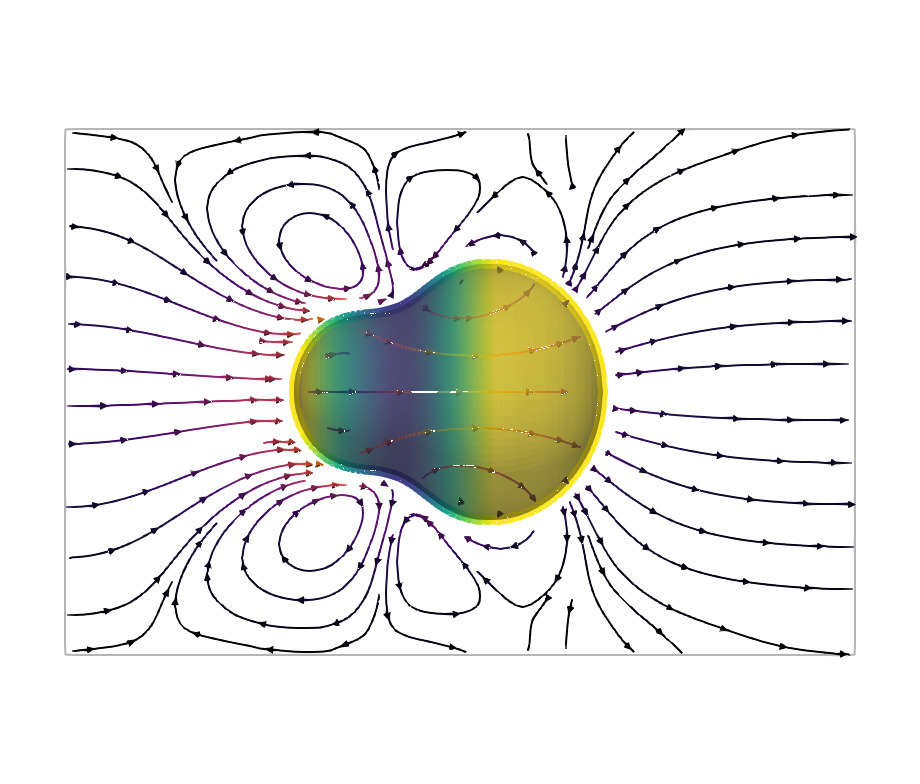}
    \vspace{-2.3cm}\newline 
    \includegraphics[height=6cm]{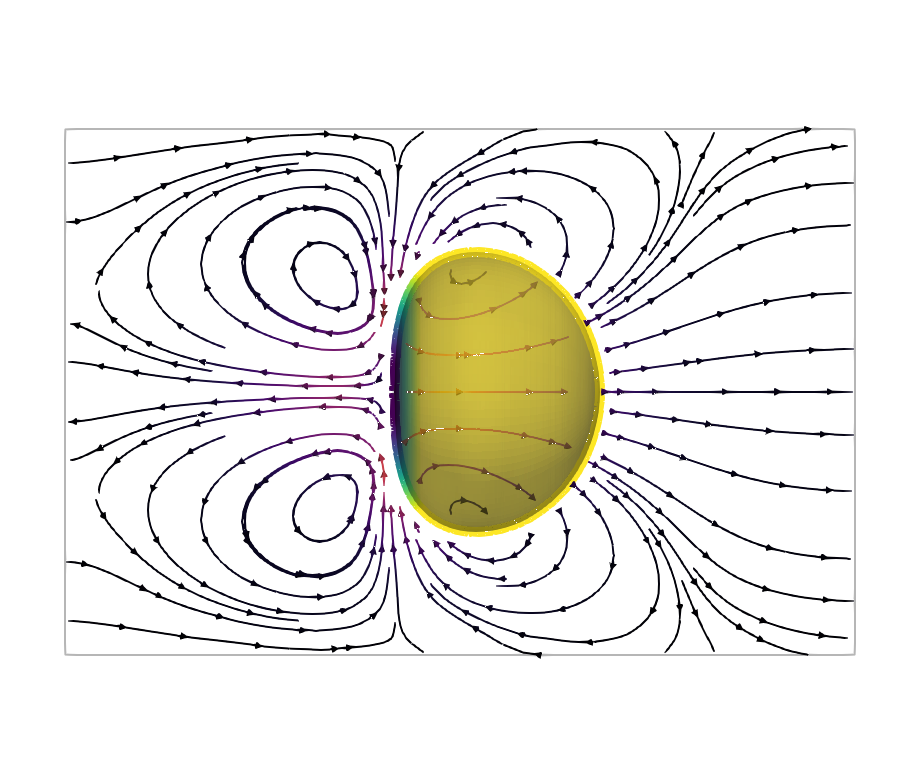}
    \end{minipage}
    \begin{minipage}{7cm}
    ~~~~~ b) Cell Division 
    \vspace{-1.0cm}\newline  
    \includegraphics[height=6cm]{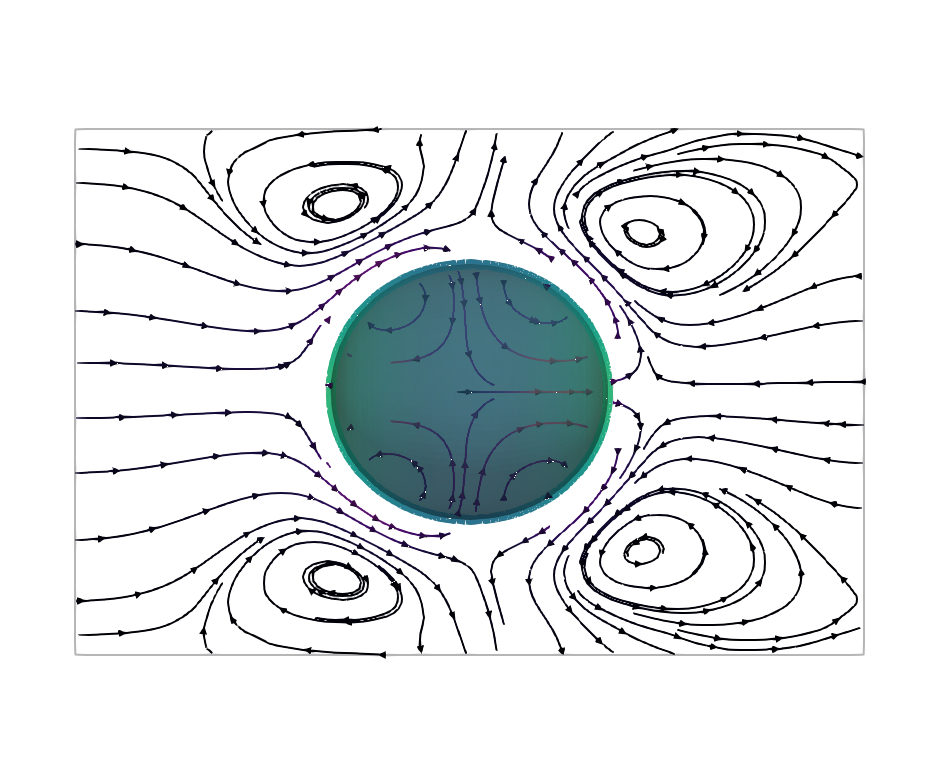}
    \vspace{-2.3cm}\newline 
    \includegraphics[height=6cm]{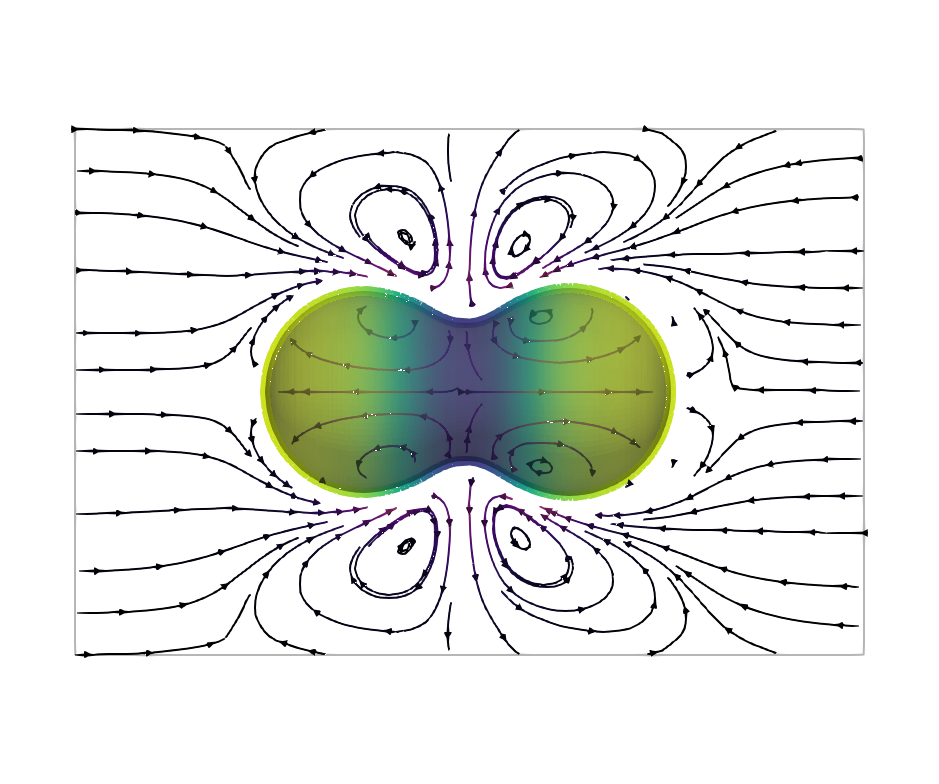}
    \vspace{-2.3cm}\newline 
    \includegraphics[height=6cm]{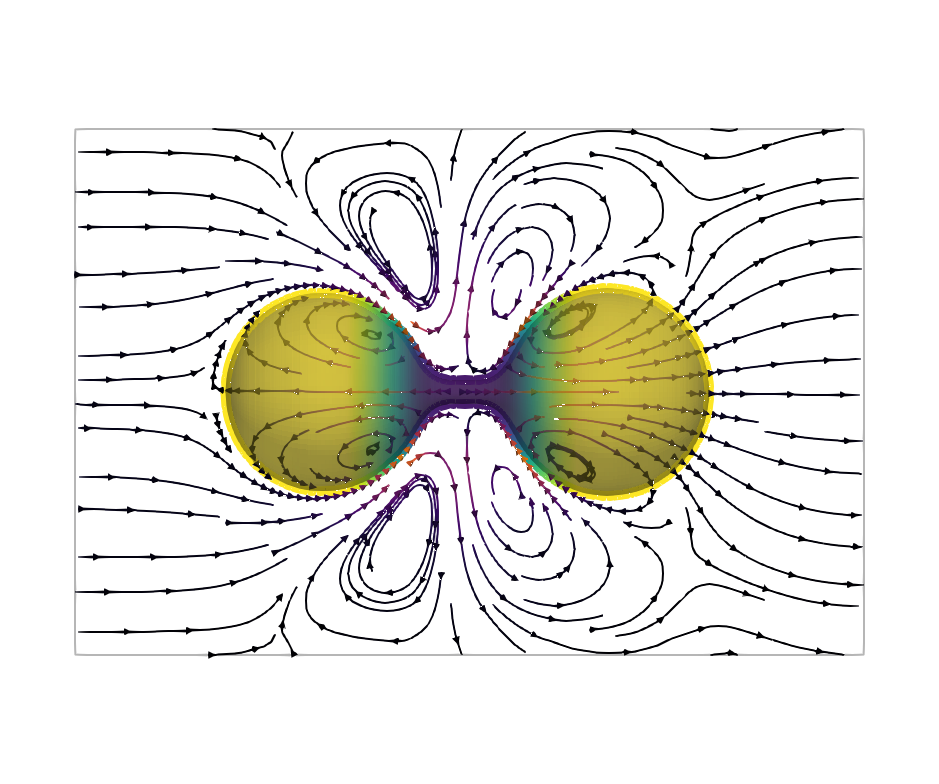}
    \vspace{-2.3cm}\newline 
    \includegraphics[height=6cm]{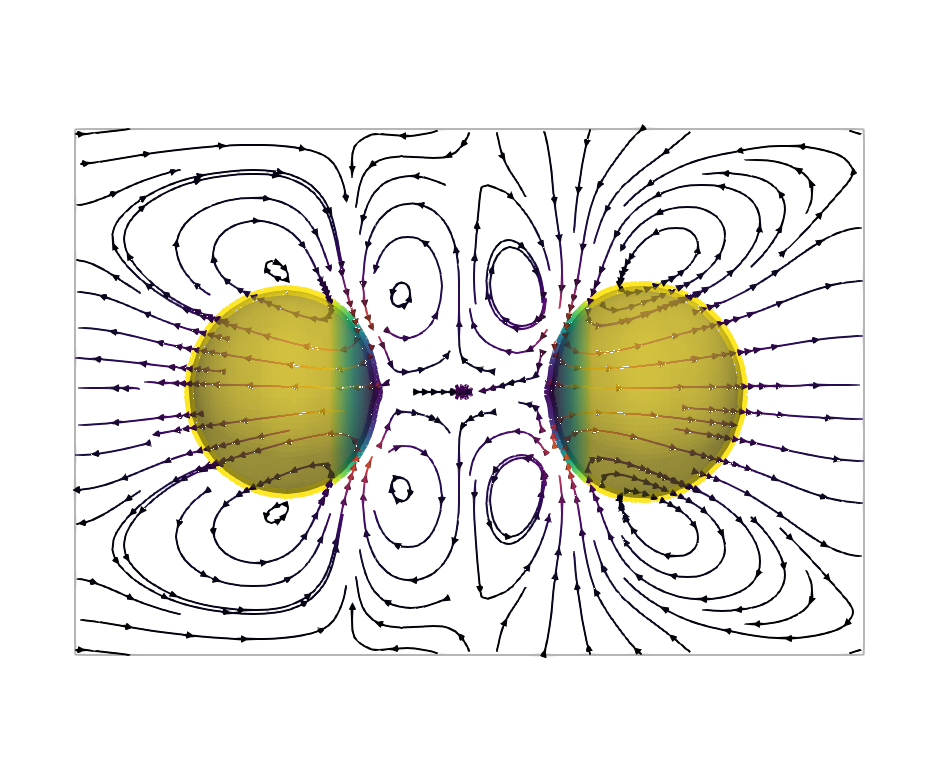}
    \end{minipage}
    \vspace{-1cm}\newline 
    \includegraphics[width=14cm]{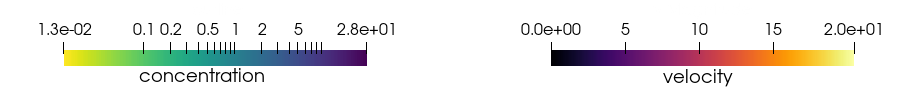}
    \caption{Pattern and shape dynamics for ring configurations in the nonlinear regime. 
    {\bf a)} For constant $\nu=\nu_0$ a strong ring pattern emerges as predicted by linear stability analysis, but slips to a pole in the nonlinear regime. Shown time points \(t=0.0, 0.4, 0.8, 1.2\).
    {\bf b)} For $c$-dependent $\nu$ (see Eq.~\eqref{eq:nu of c}) the ring pattern remains stable and goes along with a strong constriction finally leading to division of the cell. Shown time points \(t=0.0, 0.2, 0.5, 0.6\).    
    Parameters: $\gamma = 1, \Pe = 75, \nu_0=1, \eta=1.0+0.1, p\in\{0,1\}$.}
    \label{fig:division}
\end{figure}

We want to use our numerical method to go beyond that and investigate the non-linear system in this parameter regime. We initialize $c$ as a ring of higher concentration around the equator and study its self-reinforcing behavior for \(\Pe \gg \Pe_2^*\). Since in this experiment we expect higher velocities to occur and push the cell out of the domain, we subtract the mean cell velocity in the advection terms causing the cell to stay in the center of the computational domain. The shape changes are not influenced by that.
Results are depicted in Fig.~\ref{fig:division}a. Similarly to the findings in \cite{AlandWittwer_2023}, we observe that the \(l=2\) mode leads to an unstable equilibrium point that eventually turns into a more dominant polar pattern. Consistent with the results in \cite{AlandWittwer_2023}, this behavior, also termed ring-slipping, is observed whenever the activity is high enough to induce significant shape changes. 

\begin{figure}
    \centering     
    \includegraphics[width=\textwidth]{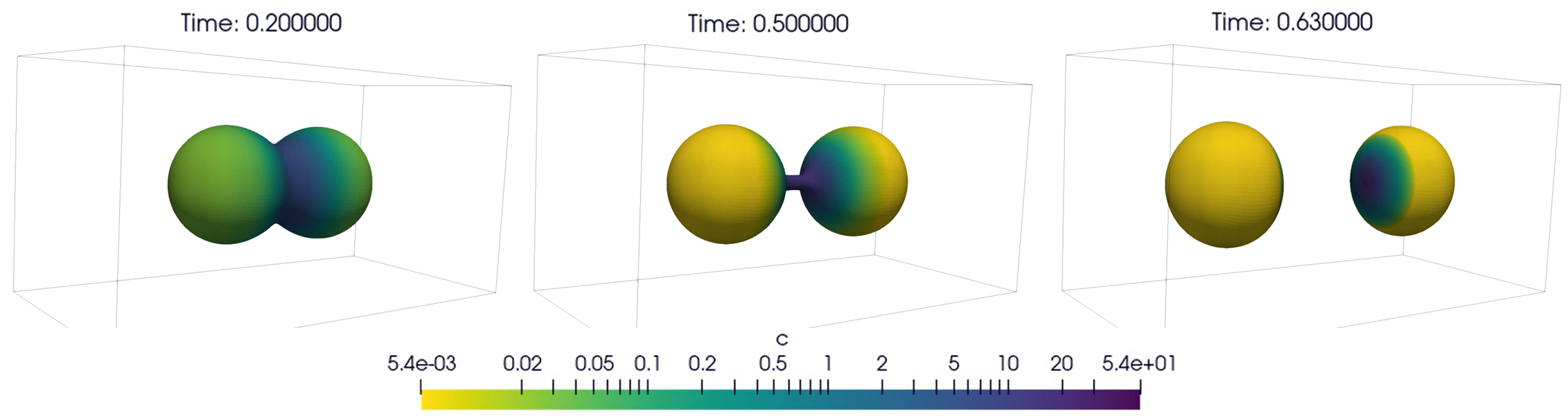}
    \caption{3D visualization of the cell division process shown in Fig.~\ref{fig:division} (right).}
    \label{fig:division 3d}
\end{figure}

One way to stabilize the contractile ring was proposed in \cite{Bonati_2022} by taking the shear-stiffening of biopolymeric networks into account. 
Correspondingly, the shear viscosity increases with the surface concentration, which was proposed to be captured by choosing the viscosity ratio
\begin{equation}
  \nu(c) = \nu_0 \frac{(1+p) c^2}{pc_0^2+c^2}
  \label{eq:nu of c}
\end{equation}
with \(\nu_0, p\geq 0\). For \(p=0\) the viscosity ratio remains constant \(\nu(c)\equiv\nu_0\) as considered before. 
For $p=1$ we observe a perfect stabilization of the ring pattern, see Fig.~\ref{fig:division}b. Correspondingly, a strong constriction builds up, finally leading to division of the cell.  After division, two daughter cells of halved volume are present. Both show opposing polar patterns such that they move apart. A 3D visualization of the process is shown in Fig.~\ref{fig:division 3d}.
To our knowledge, the present simulations are the  first simulations of the full division process based on active gel theory of the cell surface.

%% file: 6.2_extending_v.tex
\subsection{Extending a surface equation to \texorpdfstring{$\Omega$}{TEXT}} \label{sec:diffuse interface derivation}
In this subsection, we formally justify the diffuse domain formulation of the surface concentration equation \eqref{eq:phase_field:conc}. 
Therefore we extend the sharp interface equation \eqref{eq:c nondim} to the full domain \(\Omega\). To do so, we introduce the weak formulation by testing with a suitable test function $\psi$ with compact support in $\Omega\times[0,T]$. Note, that this implies that $\psi$ vanishes at the boundary of $\Omega$ and at start time $0$ and end time $T$.
This allows to write the weak formulation of Eq.~\eqref{eq:c nondim} as 
\begin{align*}
    0 &= \int_\Gamma \psi \left[ \partial_t c + \nabla_\Gamma \cdot (c \mathbf{u})- \Delta_\Gamma c \right] \dd x\\
      &= \int_\Gamma \partial_t^\bullet (c\psi) - c\partial_t^\bullet \psi + c \psi \nabla_\Gamma \cdot {\bf v} +\nabla_\Gamma c\cdot \nabla_\Gamma \psi \dd x.
\end{align*}
Using the Reynolds transport theorem on surfaces yields
\begin{align*}
    0 &= \frac{\dd}{\dd t} \int_{\Gamma(t)} \psi c \dd x +  \int_{\Gamma} -c\partial_t^\bullet \psi  +\nabla_\Gamma c\cdot \nabla_\Gamma \psi \dd x.
\end{align*}
Now the domain of integration can be extended to $\Omega \supset \Gamma$ by use of the interface Dirac-delta function $\delta_\Gamma$.  
\begin{align*}
    0 &= \frac{d}{dt} \int_{\Omega} \delta_\Gamma \psi c \dd x +  \int_{\Omega} -\delta_\Gamma c\partial_t^\bullet \psi  + \delta_\Gamma \nabla c \cdot P_\Gamma \cdot \nabla \psi \dd x.
\end{align*}
Using the ordinary Reynolds transport theorem (not on a surface) yields
\begin{align*}
    0 &=  \int_{\Omega} \partial_t^\bullet(\delta_\Gamma c) \psi + \delta_\Gamma c \psi \nabla\cdot {\bf v} + \delta_\Gamma \nabla c\cdot P_\Gamma \cdot \nabla \psi \dd x \\
    &=  \int_{\Omega} \partial_t(\delta_\Gamma c) \psi + \nabla\cdot(\delta_\Gamma c {\bf v})\psi - \nabla\cdot(\delta_\Gamma P_\Gamma \nabla c) \psi \dd x.
\end{align*}
Going back to the strong formulation gives 
\begin{align*}
    0 &=  \partial_t(\delta_\Gamma c) + \nabla\cdot(\delta_\Gamma c {\bf v}) - \nabla\cdot(\delta_\Gamma P_\Gamma \nabla c) .
\end{align*}
The asymptotic analysis of this equation without the surface projection $P_\Gamma$ shows that the concentration is constant in normal direction, ${\bf n}\cdot \nabla c=0$ for $\epsilon\rightarrow 0$. In this case $P_\Gamma\cdot \nabla c = \nabla c$, hence $P_\Gamma$ can be omitted. 
Approximating the surface delta function by $|\nabla\phi|$ yields the diffuse interface concentration equation \eqref{eq:phase_field:conc}.

%% file: 6.3_asymptotc_analysis.tex
\subsection{Asymptotic analysis} \label{sec:asymptotic}
We will argue in this section, that the solution \(\mathbf{v}(x, \varepsilon)\) of
\begin{align}
    \vert \nabla \phi \vert \mathbf{v}-\vert \nabla \phi \vert\mathbf{u}-\nabla \cdot [\vert \nabla \phi \vert( \mathbf{ n} \times \mathbf{ n}) \nabla \mathbf{v}] &= \mathbf{0} &\text{on }\Omega,\tag{\ref{eq:phase_field:v_extended}}
\end{align}
is a constant normal extension of ${\bf u}$, i.e. for \(\varepsilon\to 0\), the equations converge formally to
\begin{align}
    \mathbf{v} &= \mathbf{u}&\text{on }\Gamma,\tag{\ref{eq:sharp_interface:u=v}}\\
    (\mathbf{n}\cdot\nabla) \mathbf{v}  &= 0 &\text{on }\Gamma\tag{\ref{eq:sharp_interphase:v'=0}}.
\end{align}
To do so, we utilize matched asymptotic analysis and  follow the argumentation of \cite{RaetzVoigt_2006} and \cite{li2009_diffuse_domain}. We start by introducing a new inner coordinate system in a neighborhood of the surface \(\Gamma\), such that for any \(x\) in the neighborhood there is an unique \(s(x)\in\Gamma\) with minimal distance to \(x\). Then \(x\) can be represented as
\begin{displaymath}
  x = s(x) + r(x)\mathbf{n} = s(x) + \varepsilon z(x) \mathbf{n},
\end{displaymath}
where \(r\) is a signed distance function with \(r<0\) in \(\Omega_1\) and \(r>0\) in \(\Omega_0\). The variable $z$ is a scaled distance function defined by \(z(x)=\frac{r(x)}{\varepsilon}\).
The phase-field function can be expressed in terms of these coordinates as 
\begin{align}
    \phi
    = \frac{1}{2} \left(1-\tanh\left(\frac{r}{\sqrt{2}\varepsilon}\right)\right) 
    = \frac{1}{2} \left(1-\tanh\left(\frac{z}{\sqrt{2}}\right)\right). \label{eq:phi of z}
\end{align}

We now expand \(\mathbf{u}(x, \varepsilon)\) and \(\mathbf{v}(x, \varepsilon)\) outside the interface region in terms of the original coordinate system \(\mathbf{u} = \sum_{i=0}^\infty \varepsilon^i u^{(i)}, \mathbf{v} = \sum_{i=0}^\infty \varepsilon^i v^{(i)}\). 
This is called the outer expansion.
In the neighborhood of \(\Gamma\) we introduce for \(\mathbf{\hat u}(s, z)\) and \(\mathbf{\hat v}(s,  z)\)  the inner expansion \(\mathbf{\hat u} = \sum_{i=0}^\infty \varepsilon^i \hat u^{(i)}, \mathbf{\hat v} = \sum_{i=0}^\infty \varepsilon^i \hat v^{(i)}\) in terms of the new coordinate system. In an overlapping region, both are valid representations of the same function and thus, the following matching conditions hold
\begin{align}
   & \lim_{r\to \pm 0} u^{(0)}(s,r) &=& \lim_{z\to\pm\infty} \hat u^{(0)}(s,z)\label{eq:A:matching4}\\
   & \lim_{r\to \pm 0} v^{(0)}(s,r) &=& \lim_{z\to\pm\infty} \hat v^{(0)}(s,z)\label{eq:A:matching5}\\
   & \lim_{r\to \pm 0} \nabla v^{(0)}(s,r) \cdot \mathbf{n} &=& \lim_{z\to\pm\infty} \partial_z \hat v^{(1)}(s,z).\label{eq:A:matching6}
\end{align}
Inserting the \textbf{outer expansions} into Eq.~\eqref{eq:phase_field:v_extended} then yields
\begin{align*}
 0=\vert \nabla \phi^{(0)} \vert v^{(0)}-\vert \nabla \phi^{(0)} \vert u^{(0)}-\nabla \cdot [\vert \nabla \phi^{(0)} \vert( \mathbf{n} \times \mathbf{n}) \nabla v^{(0)}],
\end{align*}
which gives the trivial identity $0=0$ away from the interface. 

In the inner coordinate system, Eq.~\eqref{eq:phase_field:v_extended} turns to
\begin{align*}
 0 &= \vert \nabla \phi \vert \mathbf{\hat v}-\vert \nabla \phi \vert\mathbf{\hat u}-(\frac{1}{\varepsilon}\textbf{n}\partial_z + \nabla_\Gamma) \cdot [\vert \nabla \phi \vert( \mathbf{n} \times \mathbf{n}) (\frac{1}{\varepsilon}\textbf{n}\partial_z + \nabla_\Gamma) \mathbf{\hat v}] \\
 &=\vert \nabla \phi \vert \mathbf{\hat u}-\vert \nabla \phi \vert\mathbf{\hat v}+\frac{1}{\varepsilon^2} \partial_z \left(\vert \nabla \phi\vert \partial_z \mathbf{\hat v}\right) + \frac{1}{\varepsilon} \nabla_\Gamma \cdot \left(\vert \nabla \phi\vert \mathbf{n} \partial_z \mathbf{\hat v}\right),
\end{align*}
where we used ${\bf n}\cdot\nabla_\Gamma=0$ and $\partial_z {\bf n}=0$. 
By inserting the \textbf{inner expansion} and comparing powers of \(\varepsilon\), we obtain the following condition at order $1/\varepsilon^2$
\begin{align*}
    0= \partial_z \left(\vert \nabla \phi\vert \partial_z \hat v^{(0)}\right).
\end{align*}
Thus, \(\vert \nabla \phi\vert \partial_z \hat v^{(0)}\) is constant in $z$. This constant value must be zero since $|\nabla\phi|\rightarrow 0$ for $z\rightarrow\pm\infty$ (see Eq.~\eqref{eq:phi of z}).
Hence, \(\partial_z {\hat v^{(0)}}=0\). Using this result, we obtain at order $1/\varepsilon$
\begin{align*}
    0 = \partial_z \left(\vert \nabla \phi\vert \partial_z \hat v^{(1)}\right).
\end{align*}
Similarly to above, we deduce \(\partial_z \hat v^{(1)}=0\). Using matching condition \eqref{eq:A:matching6} we obtain the desired condition \eqref{eq:sharp_interphase:v'=0}, as
\begin{align*}
 (\mathbf{n}\cdot\nabla) \mathbf{v}^{(0)}  &= 0 & {\rm on~}\Gamma.
\end{align*}
Finally, we have at order $1$ 
\begin{align*}
    0=&\vert \nabla \phi \vert \hat u^{(0)}-\vert \nabla \phi \vert \hat v^{(0)}+\partial_z \left(\vert \nabla \phi\vert \partial_z \hat v^{(2)}\right) + \nabla_\Gamma \cdot \left(\vert \nabla \phi\vert \mathbf{n} \partial_z \hat v^{(1)}\right)\\
    =&\vert \nabla \phi \vert \hat u^{(0)}-\vert \nabla \phi \vert \hat v^{(0)}+\partial_z \left(\vert \nabla \phi\vert \partial_z \hat v^{(2)}\right).
\end{align*}
Integrating from \(z=-\infty\) to \(z=+\infty\) yields
\begin{align*}
    0 =& \int_{-\infty}^{+\infty} \left\{\vert \nabla \phi \vert \hat u^{(0)}-\vert \nabla \phi \vert \hat v^{(0)}+\partial_z \left(\vert \nabla \phi\vert \partial_z \hat v^{(2)}\right) \right\} \dd z\\
    =& ~\hat u^{(0)}\int_{-\infty}^{+\infty} \vert \nabla \phi \vert \dd z  - \hat v^{(0)} \int_{-\infty}^{+\infty} \vert \nabla \phi \vert \dd z + \left[ \vert \nabla \phi\vert \partial_z \hat v^{(2)}\right]_{-\infty}^{+\infty},
\end{align*} 
where we used that \(\hat v^{(0)}\) is constant in $z$ as shown above and \(\hat u^{(0)}\) is constant in $z$ since ${\bf u}$ is a continuous function (actually: $\hat u^{(0)} = {\bf u}_{|\Gamma}$). Further, from Eq.~\eqref{eq:phi of z} we conclude for $z=\pm\infty$ that  $|\nabla\phi|=0$, hence \(\left[ \vert \nabla \phi\vert \partial_z \hat v^{(2)}\right]_{-\infty}^{+\infty}=0\). 
Dividing by the (non-zero) integrals, we obtain $\hat v^0= \hat u^0$. With conditions \eqref{eq:A:matching4}-\eqref{eq:A:matching5} we conclude $v^0=  u^0$, from which we recover the desired Eq.~\eqref{eq:sharp_interface:u=v}.

%% file: 6.1_axisymmetry.tex
\subsection{Axisymmetric formulation} \label{app:axisymmetric}

As hinted before we use the rotational symmetric setup to our advantage and save computation time by solving a 2D-problem and then rotate the solution around the \(z\)-axis turning it into 3D. For this we consider cylinder coordinates \((\rho, \varphi, z)\) throughout this section. A vector \(\boldsymbol{u} \in \mathbb{R}^3\) is then represented by \(\boldsymbol{u} = u_\rho \boldsymbol{e}_\rho+ u_\varphi \boldsymbol{e}_\varphi + u_z \boldsymbol{e}_z\) and we write \(\boldsymbol{u}=(u_\rho, u_\varphi, u_z)^T\) and use that notation for all vectors and matrices following.

To reduce the complexity of the problem, we utilize that the solution will be axisymmetric, and thus constant in \(\varphi\)-direction. We therefore replace the usual 3D-Cartesian differential operators by its cylindrical equivalent without the \(\varphi\)-direction. 

We denote the unit vectors in cylindrical coordinates by \(\boldsymbol{e}_\rho\), \(\boldsymbol{e}_\varphi\) and \(\boldsymbol{e}_z\). It holds
\begin{align*}
    &\partial_\rho \boldsymbol{e}_\rho = 0, && \partial_\varphi \boldsymbol{e}_\rho = \boldsymbol{e}_\varphi, &&\partial_z \boldsymbol{e}_\rho = 0,\\
    &\partial_\rho \boldsymbol{e}_\varphi = 0, && \partial_\varphi \boldsymbol{e}_\varphi = -\boldsymbol{e}_\rho, &&\partial_z \boldsymbol{e}_\varphi = 0,\\
    &\partial_\rho \boldsymbol{e}_z = 0, && \partial_\varphi \boldsymbol{e}_z = 0, &&\partial_z \boldsymbol{e}_z = 0.
\end{align*}
In cylindrical coordinates, the gradient is given by 
\begin{displaymath}
  \nabla_R \coloneqq \left[\boldsymbol{e}_\rho \partial_\rho + \boldsymbol{e}_\varphi(\frac{1}{\rho} \partial_\varphi)  + \boldsymbol{e}_z \partial_z \right] \otimes.
\end{displaymath}
Specifically, the gradient of a scalar field \(f\) is 
\begin{displaymath}
    \nabla_R f = \partial_\rho f \boldsymbol{e}_\rho + (\frac{1}{\rho} \partial_\varphi f) \boldsymbol{e}_\varphi + \partial_z f \boldsymbol{e}_z = (\partial_\rho f, \frac{1}{\rho}\partial_\varphi f, \partial_z f)^T
\end{displaymath}
and the gradient of a vector field \(\boldsymbol u=u_\rho \boldsymbol{e}_\rho+ u_\varphi \boldsymbol{e}_\varphi + u_z \boldsymbol{e}_z\)
\begin{displaymath}
    \nabla_R \boldsymbol u =\boldsymbol{e}_\rho  \otimes(\partial_\rho \boldsymbol u )  + \boldsymbol{e}_\varphi \otimes (\frac{1}{\rho} \partial_\varphi \boldsymbol u) + \boldsymbol{e}_z \otimes (\partial_z \boldsymbol u)  =
    \begin{pmatrix}
    \partial_\rho u_\rho & \partial_\rho u_\varphi & \partial_\rho u_z \\
    \frac{1}{\rho}(\partial_\varphi u_\rho - u_\varphi) & \frac{1}{\rho}(\partial_\varphi u_\varphi + u_\rho) & \frac{1}{\rho} \partial_\varphi u_z  \\
    \partial_z u_\rho & \partial_z u_\varphi & \partial_z u_z
    \end{pmatrix}.
\end{displaymath}
The divergence is defined in a similar fashion as the inner product
\begin{displaymath}
    \nabla_R \cdot \coloneqq \left[\boldsymbol{e}_\rho \partial_\rho + \boldsymbol{e}_\varphi (\frac{1}{\rho} \partial_\varphi) + \boldsymbol{e}_z \partial_z \right] \cdot,
\end{displaymath}
thus the divergence of a vector field \(\boldsymbol u\) is given by
\begin{displaymath}
    \nabla_R \cdot \boldsymbol{u} = \partial_\rho u_\rho + \frac{1}{\rho} (\partial_\varphi u_\varphi + u_\rho) + \partial_z u_z. 
\end{displaymath}
We observe for a tensor \(A\in\mathbb{R}^{n\times n}\)
\begin{align*}
   \nabla_R \cdot A &= \nabla_R \cdot 
    \begin{pmatrix}
    a_{\rho,\rho} & a_{\rho,\varphi} & a_{\rho,z}\\
    a_{\varphi,\rho} & a_{\varphi,\varphi} &  a_{\varphi,z}\\
    a_{z,\rho} & a_{z,\varphi} & a_{z,z}
    \end{pmatrix}\\
    &= \sum_{k,l=\rho, \varphi, z} \boldsymbol{e}_\rho \cdot \partial_\rho (a_{k,l} \boldsymbol{e}_k\otimes \boldsymbol{e}_l) 
    + \sum_{k,l=\rho, \varphi, z} \boldsymbol{e}_\varphi \cdot \frac{1}{\rho} \partial_\varphi (a_{k,l} \boldsymbol{e}_k\otimes \boldsymbol{e}_l) 
    + \sum_{k,l=\rho, \varphi, z} \boldsymbol{e}_z \cdot \partial_z (a_{k,l} \boldsymbol{e}_k\otimes \boldsymbol{e}_l)\\
    &= \sum_{l=\rho, \varphi, z} (\partial_\rho a_{\rho,l}) \boldsymbol{e}_l 
    + \frac{1}{\rho} \sum_{l=\rho, \varphi, z} (\partial_\varphi a_{\varphi,l} + a_{\rho,l}) \boldsymbol{e}_l +a_{\varphi,l} (\partial_\varphi \boldsymbol{e}_l)
    + \sum_{l=\rho, \varphi, z} (\partial_z a_{z,l}) \boldsymbol{e}_l\\
   &= \sum_{\substack{k=\rho, z \\ l=\rho, \varphi, z}} (\partial_k a_{k,l}) \boldsymbol{e}_l 
    + \frac{1}{\rho} \sum_{l=\rho, \varphi, z} (\partial_\varphi a_{\varphi,l} + a_{\rho,l}) \boldsymbol{e}_l + \frac{1}{\rho} a_{\varphi,\rho} \boldsymbol{e}_\varphi -\frac{1}{\rho} a_{\varphi, \varphi} \boldsymbol{e}_\rho.
\end{align*}
Furthermore, the normal vector on the surface \(\Gamma\) is of the form \(\boldsymbol n = n_\rho \boldsymbol{e}_\rho+n_z\boldsymbol{e}_z\) and 
\begin{displaymath}
    P_\Gamma = \begin{pmatrix}1-n_\rho n_\rho & 0& -n_\rho n_z\\0&1&0\\-n_\rho n_z&0&1-n_zn_z \end{pmatrix} 
    = \begin{pmatrix}p_{\rho,\rho} & 0& p_{\rho,z}\\0&p_{\varphi,\varphi}&0\\p_{\rho,z}&0&p_{z,z} \end{pmatrix} .
\end{displaymath}
Now we utilize that our system is constant in azimuthal direction, in particular all first derivatives in \(\varphi\)-direction vanish and \(u_\varphi=0\). With that, the incompressibility condition \eqref{eq:NS 2 nondim} turns to 
\begin{displaymath}
    \nabla_R\cdot \boldsymbol{u} = \left[\partial_\rho \boldsymbol{e}_\rho + (\frac{1}{\rho} \partial_\varphi) \boldsymbol{e}_\varphi + \partial_z \boldsymbol{e}_z\right] \cdot \boldsymbol u = \partial_\rho u_\rho + \frac{1}{\rho}\left(\partial_\varphi u_\varphi + u_\rho\right) +\partial_z u_z \overset{u_\varphi=0}{=} \frac{1}{\rho}\partial_\rho\left(\rho u_\rho \right) + \partial_z u_z.
\end{displaymath}
For the conservation of momentum, Eq.~\eqref{eq:NS 1 phase-field},   
\begin{align}
    -\nabla \cdot \left[\eta(\nabla \mathbf{u} + \nabla \mathbf{u}^T )\right] + \nabla p =&
    \nabla \cdot \left[\delta_\Gamma(1-\nu)P_\Gamma{\tilde\nabla}_\Gamma \cdot \mathbf{u} +\nu \delta_\Gamma P_\Gamma(\nabla \mathbf{u} + \nabla \mathbf{u}^T)P_\Gamma  \right] \nonumber\\
    &+\Pe\delta_\Gamma \left(3\sqrt{2} \sigma(c) \mu\nabla\phi + P_\Gamma { \nabla} \sigma(c)\right).
\end{align}
Note, that for shorter notation we write here $\delta_\Gamma$ instead of the $|\nabla\phi|$. 
Now formulating this in the described axisymmetric setting, yields for the first order derivatives
\begin{align*}
  \nabla_R p &= (\partial_\rho p) \boldsymbol{e}_\rho + (\partial_z p) \boldsymbol{e}_z\\
  \nabla_R \phi &= (\partial_\rho \phi) \boldsymbol{e}_\rho + (\partial_z \phi) \boldsymbol{e}_z\\
  P_\Gamma \nabla\sigma(c) &= \sigma'(c) P_\Gamma \nabla_R c = \sigma'(c) P_\Gamma ((\partial_\rho c) \boldsymbol{e}_\rho + (\partial_z c) \boldsymbol{e}_z)
\end{align*}
and for the second order derivatives
\begin{align*}
    &\nabla_R \cdot \left[\eta(\nabla_R \mathbf{u} + \nabla_R \mathbf{u}^T )\right] 
    &=& \nabla_R \cdot \eta
    \begin{pmatrix}
    2\partial_\rho u_\rho & 0 & \partial_z u_\rho+\partial_\rho u_z\\
    0 & \frac{2}{\rho} u_\rho & 0 \\
    \partial_\rho u_z+\partial_z u_\rho & 0 & 2\partial_z u_z
    \end{pmatrix}\\
    &&=& \sum_{k,l=\rho, z} \partial_k(\eta \partial_l u_k+\eta\partial_k u_l) \boldsymbol{e}_l 
    + \frac{\eta}{\rho} \sum_{l=\rho, z} (\partial_l u_\rho+\partial_\rho u_l) \boldsymbol{e}_l -\frac{2\eta}{\rho^2} u_\rho \boldsymbol{e}_\rho,
    \\
    &\nabla_R \cdot \left[\nu\delta_\Gamma P_\Gamma(\nabla_R \mathbf{u} + \nabla_R \mathbf{u}^T )P_\Gamma\right] 
    &=& \nabla_R \cdot \nu \delta_\Gamma
   \left( \sum_{\substack{i,j=\rho, z\\k,l=\rho, z}} p_{k,i}(\partial_i u_j + \partial_j u_i)p_{j,l} \boldsymbol{e}_k\otimes \boldsymbol{e}_l + p_{\varphi,\varphi}\frac{2}{\rho} u_\rho p_{\varphi,\varphi} \boldsymbol{e}_\varphi\otimes \boldsymbol{e}_\varphi\right)\\
   &&=& \sum_{k,l=\rho, z} \partial_k
   \left(\nu \delta_\Gamma \sum_{i,j=\rho, z} p_{k,i}(\partial_i u_j + \partial_j u_i)p_{j,l}\right) \boldsymbol{e}_l \\
    &&&+ \frac{\nu \delta_\Gamma}{\rho} \sum_{l=\rho, z} \left(\sum_{i,j=\rho, z} p_{\rho,i}(\partial_i u_j + \partial_j u_i)p_{j,l}\right) \boldsymbol{e}_l -\frac{2\nu\delta_\Gamma}{\rho^2} u_\rho \boldsymbol{e}_\rho,
    \\
    &\nabla_R\cdot\left[(1-\nu)\delta_\Gamma {\tilde\nabla}_{\Gamma,R} \cdot \boldsymbol{u} P_\Gamma\right]
    &=&\nabla_R\cdot (1-\nu)\delta_\Gamma \left(\sum_{i,j=\rho,z} p_{i,j}\partial_i u_j + p_{\varphi,\varphi}\frac{1}{\rho}u_\rho\right) P_\Gamma\\
    &&=&\sum_{k,l=\rho, z} \partial_k
   \left((1-\nu)\delta_\Gamma \left(\sum_{i,j=\rho,z} p_{i,j}\partial_i u_j + \frac{u_\rho}{\rho}\right) p_{k,l}\right) \boldsymbol{e}_l \\
    &&&+ \frac{(1-\nu) \delta_\Gamma}{\rho} \sum_{l=\rho, z} \left(\sum_{i,j=\rho,z} p_{i,j}\partial_i u_j  + \frac{u_\rho}{\rho} \right)p_{\rho,l} \boldsymbol{e}_l \\
    &&&-\frac{(1-\nu)\delta_\Gamma}{\rho} \left(\sum_{i,j=\rho,z} p_{i,j}\partial_i u_j  + \frac{u_\rho}{\rho} \right) \boldsymbol{e}_\rho.
\end{align*}

Finally, for the concentration equation \eqref{eq:phase_field:conc} the application of rotational operators $\nabla_R$ and $\nabla_R\cdot$ is straightforward. 
Similarly, for the Cahn-Hilliard equation, $\Delta \phi$ is replaced by $\nabla_R\cdot \nabla_R \phi$, which is necessary for the chemical potential to approximate the correct 3-dimensional curvature. However, we refrain from using axisymmetric operators for $\Delta\mu$  to improve conservation of cell volume. Using rotational symmetry at this point, would effectively increase the ratio between the extracellular and the intracellular fluid volume, increasing the undesired dissolution of a part of the cell into the surrounding fluid domain. 
Note that, since the primary purpose of the phase-field is to track the cellular interface, rotationally symmetric operators are not necessary for $\Delta \mu$.
